\documentclass[letterpaper, 10 pt, conference]{ieeeconf}  

\usepackage[vlined,ruled]{algorithm2e}

\usepackage{caption}
\usepackage[normalem]{ulem}
\usepackage{multirow}
\usepackage{makecell}
\usepackage{mathtools}
\usepackage{microtype}
\usepackage{comment}
\usepackage{graphicx}
\usepackage{subfig}

\usepackage{color}
\usepackage{epsf}
\makeatletter
\usepackage{authblk}
\renewcommand\AB@affilsepx{ --- \protect\Affilfont}
\newif\if@restonecol
\makeatother

\usepackage{tikz}
\usetikzlibrary{calc,shapes.multipart,chains,arrows}

\usepackage{booktabs}

\usepackage{flushend}

\begin{document}

\title{Performance Impact of Memory Channels on Sparse and Irregular Algorithms}


\bstctlcite{IEEEexample:BSTcontrol}



\author[1,2]{Oded~Green}
\author[2]{James~Fox}
\author[2]{Jeffrey~Young}
\author[2]{Jun~Shirako}
\author[3]{David~Bader}
\affil[1]{NVIDIA Corporation}
\affil[2]{Georgia Institute of Technology}
\affil[3]{New Jersey Institute of Technology}
\maketitle

\thispagestyle{plain}
\pagestyle{plain}

\newcommand{\BC}{Betweenness Centrality\xspace}


\newcommand{\CC}{Connected Components\xspace}
\newcommand{\PR}{PageRank\xspace}  
\newcommand{\SSSP}{Single-Source Shortest Path\xspace} 
\newcommand{\BFS}{BFS\xspace}  

\newcommand{\TC}{Triangle Counting\xspace} 

\newcommand{\NVG}{nvGRAPH\xspace} 

\begin{abstract}

Graph processing is typically considered to be a memory-bound rather than compute-bound problem. One common line of thought is that more available memory bandwidth corresponds to better graph processing performance. However, in this work we demonstrate that the key factor in the utilization of the memory system for graph algorithms is not necessarily the raw bandwidth or even the latency of memory requests. Instead, we show that performance is proportional to the number of memory channels available to handle small data transfers with limited spatial locality.

Using several widely used graph frameworks, including Gunrock (on the GPU) and GAPBS \& Ligra (for CPUs), we evaluate key graph analytics kernels using two unique memory hierarchies, DDR-based and HBM/MCDRAM. Our results show that the differences in the peak bandwidths of several Pascal-generation GPU memory subsystems aren't reflected in the performance of various analytics. Furthermore, our experiments on CPU and Xeon Phi systems demonstrate that the number of memory channels utilized can be a decisive factor in performance across several different applications. For CPU systems with smaller thread counts, the memory channels can be underutilized while systems with high thread counts can oversaturate the memory subsystem, which leads to limited performance. Finally, we model the potential performance improvements of adding more memory channels with narrower access widths than are found in current platforms, and we analyze performance trade-offs for the two most prominent types of memory accesses found in graph algorithms, streaming and random accesses.

\end{abstract}



\section{Introduction}
\label{sec:intro}
 
Graph processing is usually considered memory-bound due to the irregular and data-dependent nature of most graph problems, which leads to many irregular memory accesses. It is also commonly believed that these algorithms contain a mix of bandwidth- and latency-bound operations. For modern shared memory systems that handle these types of applications, there has been an explosion in new memory technologies as well as the amount of parallelism for computation and memory accesses. For example, a Power 9 system might have multiple processors in a single node, each with up to 24 cores and 96 threads. Similarly, Intel's Knights Landing (KNL)  processor can have up to 68 cores and 272 threads, and the latest generation of NVIDIA's Volta GPUs can support up to 5120 threads on 80 streaming multi-processors. This increase in thread parallelism has been combined with faster and more complex memory technologies and standards in the last decade, including DDR4, DDR5 (expected to be released in 2020), GDDR5, Hybrid Memory Cube (HMC), and High Bandwidth Memory 2.0 (HBM2).

Graph algorithms are typically latency-bound if there is not enough parallelism to saturate the memory subsystem. However, the growth in parallelism for shared-memory systems brings into question whether graph algorithms are still primarily latency-bound. Getting peak or near-peak bandwidth of current memory subsystems requires highly parallel applications as well as good spatial and temporal memory reuse. The lack of spatial locality in sparse and irregular algorithms means that prefetched cache lines have poor data reuse and that the effective bandwidth is fairly low. The introduction of high-bandwidth memories like HBM and HMC have not yet closed this inefficiency gap for latency-sensitive accesses \cite{Radulovic:2015:stacked_dram_memsys15}. However, high-bandwidth memory has introduced a higher number of memory channels for stacked DRAMs, which can process a higher number of outstanding memory transactions. For this work, we define a \textbf{memory channel} as a logically grouped set of DRAM DIMMs for traditional memory systems (or a logical grouping of TSVs within a 3D stacked memory like HBM or HMC). Each channel is able to service and output data requests independently. 

In this work, we analyze the performance and scalability of graph algorithms and their memory access characteristics as the number of threads is increased on a wide range of CPU and GPU systems. The fact that these systems have such a high thread count is crucial as the phenomena of over-saturating the memory subsystem is not visible for smaller thread counts (even on new CPU systems with tens of cores). 

\subsection*{Contributions}
In this paper we challenge the commonly held notion that graph algorithms are either memory bandwidth- or latency-bound. Instead we show that the performance of these algorithms is dependent on the number of memory channels in the memory subsystem. Using a wide range of CPU and GPUs with a mix of DDR4, GDDR5, HBM2, and Multi-Channel DRAM (MCDRAM), we benchmark numerous analytics from highly optimized frameworks across a wide range of graphs with different properties. Contrary to past research that focused on analyzing only large thread counts, our analysis includes threads counts at different ranges, which allows us to find the point at which the memory subsystem is stressed. 

\textbf{Key findings}:  For the Intel KNL processor, which has both DDR4 and MCDRAM memories, MCDRAM starts to outperform DDR4 at around 64 threads. 
Neither DDR nor MCDRAM are saturated with less than 64 threads, and in this regime there is little performance difference due to the similar latencies of the memories.

For the NVIDIA GPU, we compare an HBM2-based GPU system with several GDDR5 and GDDR5X GPUs. While the GDDR5-based GPU's specifications suggest that peak available bandwidth has a large role to play, the HBM2 GPU outperforms them by over a factor of 2X even when the bandwidth relative to a GDDR5 device is less than 1.33X. Furthermore, there is little variance in performance across the GDDR5-based GPUs. We show that the performance is correlated to the number of memory channels available to the system - {\textbf{a number that is rarely reported when conducting performance analysis}}.

Lastly, we present a performance model that projects the performance impact of added, narrower memory channels. Our model does not focus on how such a subsystem should be created, but rather evaluates performance trade-offs from an application's point of view. 
This experiments dives into \PR and characterizes its performance with respect to streaming and random accesses in conjunction with empirical results from earlier experiments. Our findings show that increasing the number of narrower channels can greatly benefit the overall performance of sparse analytics.

\section{Related Work}
\label{sec:related}

\textbf{Applications:}
Recently, Beamer \emph{et al.} \cite{beamerlocality} showed that memory bandwidth is not fully utilized for several widely used graph kernels on a multi-threaded processor. We extend this analysis and show that for the same types of applications analyzed in Beamer \emph{et al.} \cite{beamerlocality} it is quite likely that the fairly small number of thread counts (32 threads with 16 cores) was {\bf not large enough} to saturate the memory subsystem, which in part led to underutilization of available memory bandwidth.

Xu et. al. \cite{xu2014graph} profiled a spectrum of graph applications on the GPU and found that long memory latencies as a result of high L1 cache miss rates tended to be the biggest performance bottleneck. 


Peng et. al. \cite{peng2017exploring} analyzed the performance of different benchmarks for large numbers of threads on MCDRAM vs. DRAM on the KNL system. They concluded that sequential access, bandwidth-bound applications benefit from the MCDRAM but random access benchmarks are latency-bound and perform better on DRAM. It should be noted their benchmarks were limited to random access benchmarks or BFS, as opposed to more involved sparse and irregular algorithms.

Algorithmic optimizations have been developed to improve the spatial locality of graph analytics kernels by reducing the number of cache misses \cite{beamer2017reducing,zhou2017design,zhou2015optimizing,buono2016optimizing}, but these approaches are typically application-dependent.
\label{sec:memory}

\textbf{Architectural Approaches:}
From a memory device perspective, several researchers have investigated the characteristics of the memory subsystem, but there has not yet been an evaluation that compares graph workloads across CPU, GPU, and KNL devices. We present related work that evaluates novel architecture approaches, including \textit{Half-DRAM} \cite{zhang2014half}, fine-grained DRAM \textit{FGDRAM} for GPU systems \cite{o2017fine}, and near-memory systems like the Emu Chick \cite{Dysart:2016:emuchick}.

Brunvand et. al. \cite{brunvand2014graphics} note that the number of memory channels and clock rate determine peak memory bandwidth and correspond to the level of achievable concurrency while bank counts, access patterns, and the sophistication of memory controllers are other factors which determine the actual bandwidth achieved. 

Generally speaking, data bandwidth is given by $B = W * F$, where $W$ is data width and $F$ is data frequency. Zhang et. al. \cite{zhang2014half} propose \textit{Half-DRAM} as a way to toggle $W$ to match $B$ and to induce narrower rows, while preserving bandwidth and improving power consumption. Their implementation decouples rows into half-rows within banks, which reduces row activation power and enables a doubling of memory-level parallelism.  

Similarly, the Emu Chick architecture \cite{Dysart:2016:emuchick} divides a DRAM channel into multiple "Narrow-Channel DIMMs" (NCDIMMs) which allow for more fine-grained access for irregular applications. Recent tests of BFS and small-scale graph analytics applications \cite{hein:2019:emu_chick_topc} shows promise for the NCDIMM approach with stable performance for different real-world sparse matrix multiply inputs and comparable performance with x86 platforms for BFS on balanced (Erd\"os-R\'enyi) graphs. However, performance of graph analytics on the Emu Chick is currently limited not by the memory subsystem but by data layout and workload imbalance issues that create thread migration hotspots across the Chick's distributed nodes. 

Fine-grained DRAM or \textit{FGDRAM}, as proposed by O'Connor et. al. \cite{o2017fine}, proposes the most novel approach to scaling memory bandwidth and reducing the energy contributions of memory accesses. This work rethinks the design of stacked memory by partitioning the DRAM die into smaller independent units with dedicated channels (64 channels versus 16 on current HBM) and reduced effective row size. Results on a GPU simulator show up to several factors of improvement on irregular benchmarks, due to increased concurrency for handling requests and higher row activation rates. Our characterization in Section \ref{sec:model} looks to evaluate whether this "fine-grained" approach with more channels would lead to better performance for real-world analytics test cases. 

{\it To the best of our knowledge, this work is the first to connect previous related architectural work on finer-grained memory channels with a characterization of sparse applications for current-generation CPU and GPU architectures.}
%



\section{Results - CPU and KNL}
\label{sec:results-cpu}
\begin{table*}[t]

\tiny
\centering

\caption{GPU and CPU systems used in experiments. GPUs are PCIe-based and are Pascal generation except for the V100 Volta GPU. FLOP rate for the GPUs is for single-precision. }

\begin{tabular}{|c|c|c|c|c|c|c|c|c|} \hline 
Architecture & Processor & Frequency & Cores (threads) & LL-Cache & DRAM Size & DRAM Type & Latency (ns) \cite{peng2017exploring} & Max Channels \\ \hline \hline
Knight's Landing & Intel Xeon Phi & 1.3 GHz & 64 (256) & 32MB & 16 GB/ 96 GB& MCDRAM/DDR4 & 154/130 & 32/6 \\ \hline
\end{tabular}

\vspace{0.5cm}
\begin{tabular}{|c|c|c|c|c|c|c|c|c|c|c|c|c|} \hline 
Architecture & Processor & Base & SMs & Total & L2 Size & DRAM & Size & Mem Clock & Bandwidth & Bus width & Memory & SP FLOPs \\ 
 & & Clock & & SPs & (MB) & Type & (GB) & (MT/s) & (GB/s) & (bits) & Channels & (TFLOP/s) \\ \hline \hline

GPU-CUDA     & P100  & 1126 & 56 & 3854 & 4 & HBM2 & 16 GB & 1406 & 720 & 4096 & 32			& 9.5 \\  \hline 
GPU-CUDA     & P40  & 1303 & 30 & 3840 & 3 & GDDR5 & 24 GB & 7200 & 345 & 384 & 12			& 10 \\  \hline 
GPU-CUDA     & Titan Xp & 1480 & 30 & 3840 & 3 & GDDR5X & 12 GB & 11410 & 547.7 & 384 & 12	& 10.1 \\  \hline 
GPU-CUDA     & Titan X & 1417 & 28 & 3584 & 3 & GDDR5X & 12 GB & 10000 & 480 & 384 & 12 		& 11.3 \\  \hline \hline

GPU-CUDA     & V100  & 1370 & 80 & 5120 & 4 & HBM2 & 16 GB & 1750 & 9000 & 4096 & 32			& 14 \\  \hline

\end{tabular}

\label{tab:gpu-cpu-systems}
\end{table*}

Our CPU experiments are conducted on an Intel Knights Landing (KNL) system with 64 cores (256 hardware threads), detailed in Table \ref{tab:gpu-cpu-systems}. While the KNL line of processors is no longer in production, experiments on the system are included because: 1) KNL supports a large number of threads for a single processor, 2) It has two different memory subsystems on the same board, and 3) it is one of the few CPU processors to have HBM-like memory. New Intel processors such as Skylake and Cascade Lake are less interesting because they have fewer cores and fewer memory channels overall.

\textbf{Benchmarks:} For our benchmarks we use the GAP Benchmark Suite \cite{beamer2015gap} (GAP-BS for short) and the Ligra \cite{shun2013ligra} framework with the following applications : \BFS, \BC, \TC, and \PR. 
In some cases, for the sake of brevity we only present results for a subset of these applications.
\BFS and \BC  run-times are averaged from the top 100 degree vertices. Results for \TC and \PR are averaged over 5 rounds of running the same analytic. For GAP-BS, we report execution for Direction-Optimizing \BFS \cite{beamer2012direction}. The \BC algorithm in GAP-BS uses the standard top-down \BFS. As such we are able to analyze both types of BFS traversals within GAP-BS.
For experiments on the KNL, GAP-BS is compiled with GNU \textit{gcc} with OpenMP support, and Ligra is compiled with Intel's \textit{icpc} compiler using the $-O3$ flag and Cilk for multi-threading. Ligra requires Intel's compiler as it is Cilk based.


\textbf{Profiling setup:} For detailed runtime profiling, we rely on Intel Performance Counter Monitor (PCM) \cite{willhalm2012intel} and Intel VTune \cite{malladi2009using}. We use PCM's hardware counters for read/write traffic at the memory controllers to get total DRAM memory traffic. We have not been able to find corresponding counters for MCDRAM, so we use VTune, in a separate profiling session, to verify that last-level cache misses on the KNL are essentially the same when running with either memory subsystem.








\begin{table}[t]
\begin{center}
\tiny

\caption{Networks used in our experiments, sorted by $|V|$ (number of vertices). $|E|$ is number of undirected edges.}

\begin{tabular}{|l|c|c|} \hline
Name & $|V|$ & $|E|$ \\  \hline \hline
email-Enron & 37K & 184K \\ \hline
soc-Slashdot0902 & 82K & 504K \\ \hline
dblp & 317K & 1M \\ \hline
amazon & 335K & 926K \\ \hline
amazon0601 & 400K & 2.4M \\ \hline
kron\_g500-logn21 & 1.5M & 91M \\ \hline
skitter & 1.7M & 11M \\ \hline
orkut & 3M & 117M \\ \hline

\end{tabular}
\begin{tabular}{|l|c|c|} \hline
Name & $|V|$ & $|E|$ \\  \hline \hline
wikipedia-2007 & 3.5M & 42M \\ \hline
cit-Patents & 3.8M & 16M \\ \hline
livejournal & 4M & 35M  \\ \hline
cage15 & 5.2M & 52M \\ \hline
indochina-2004 & 7.4M & 153M \\ \hline
wb-edu & 9.5M & 48M \\ \hline
uk-2002 & 18M & 268M \\ \hline
europe\_osm & 51M & 54M \\ \hline

\end{tabular}

\label{tab:graphs}

\end{center}
\end{table}

\textbf{Inputs:} The graphs used in the experiments for both the CPU and GPU are taken from SNAP \cite{Snap-Stanford} and the Florida Matrix Market \cite{florida-matrix} repositories, with specific details listed in Table \ref{tab:graphs}. All edges are as treated as undirected. Also, all self-loops, duplicate edges, and singleton vertices are removed in a preprocessing phase. Lastly, vertex IDs are randomized to reduce bias, as some of these graphs may have been previously preprocessed to improve data locality.

\paragraph*{\textbf{System Configuration}}
The KNL system has both 16GB of MCDRAM and 96GB of DDR4 memory, with options to use one or the other in isolation. Thus, we are able to benchmark an algorithm with the exact same input for two different memory configurations. Experiments on the KNL system are conducted on a wide range of graphs, but we find that the most interesting results occur for graphs with larger vertex sets. Specifically, KNL has a fairly large last level cache (LLC) size of 32MB. As most of the algorithms we test in this paper store data on a per-vertex basis, the related arrays are typically cacheable such that there are fewer memory requests to the memory subsystem. However, when a vertex set exceeds 4M vertices, it is clear that the memory system is handling a large number of accesses such that it is ``over-utilized''. This is also apparent in the GPUs used in our experiments, but GPUs typically have smaller LLCs that are shared by more threads, so we can distinguish this behavior for smaller vertex sets.

\subsection{MCDRAM versus DRAM on the KNL}
The following experiment highlights the difference between the smaller 3D-stacked MCDRAM, an HBM analogue, and more traditional DDR memory. Specifically, we show that saturating the memory subsystem on many existing servers is challenging due to the relatively smaller number of threads that can be executed on these systems. The KNL server used in this experiment allows us to scale the number of threads to a much larger thread count than in a typical CPU system and allows us to saturate the system in a controlled manner. We toggle between using either DDR or MCDRAM using \textit{numactl}. The KNL system is also configurable in NUMA affinity mode (``cluster mode''). For our experiments we use the default ``quadrant'' mode.


\paragraph*{\bf Experiment setup}
We scale the number of threads from 1 thread up to 256 threads by powers of two. Recall that the system has 64 cores, so scaling beyond 64 threads entails having multiple threads on a core. To more accurately model sequential performance, Cilk is used with single-threaded execution of Ligra. The purpose of this experiment is to show the saturation of memory channels with a large number of requests, especially as thread-level parallelism is increased. We perform this scaling for MCDRAM separately. 

\begin{figure*}[t]
\centering

\subfloat[Breadth First Search using Direction Optimization \cite{beamer2012direction}.]{
 \includegraphics[width=0.66\columnwidth]{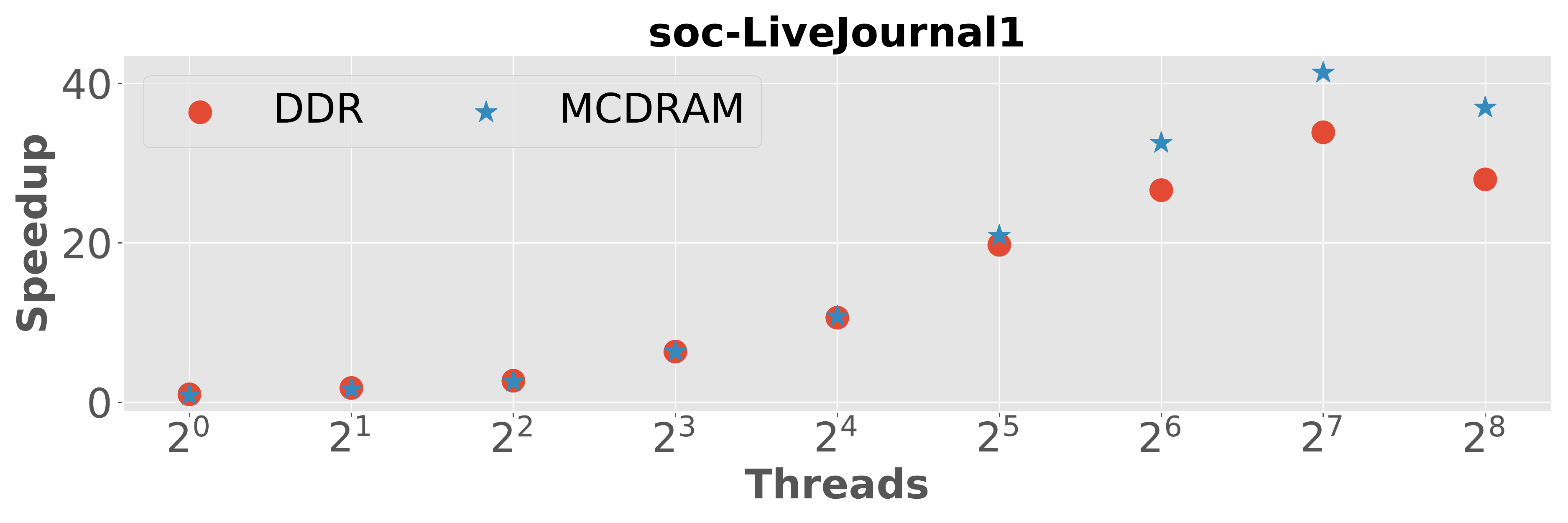}
 \includegraphics[width=0.66\columnwidth]{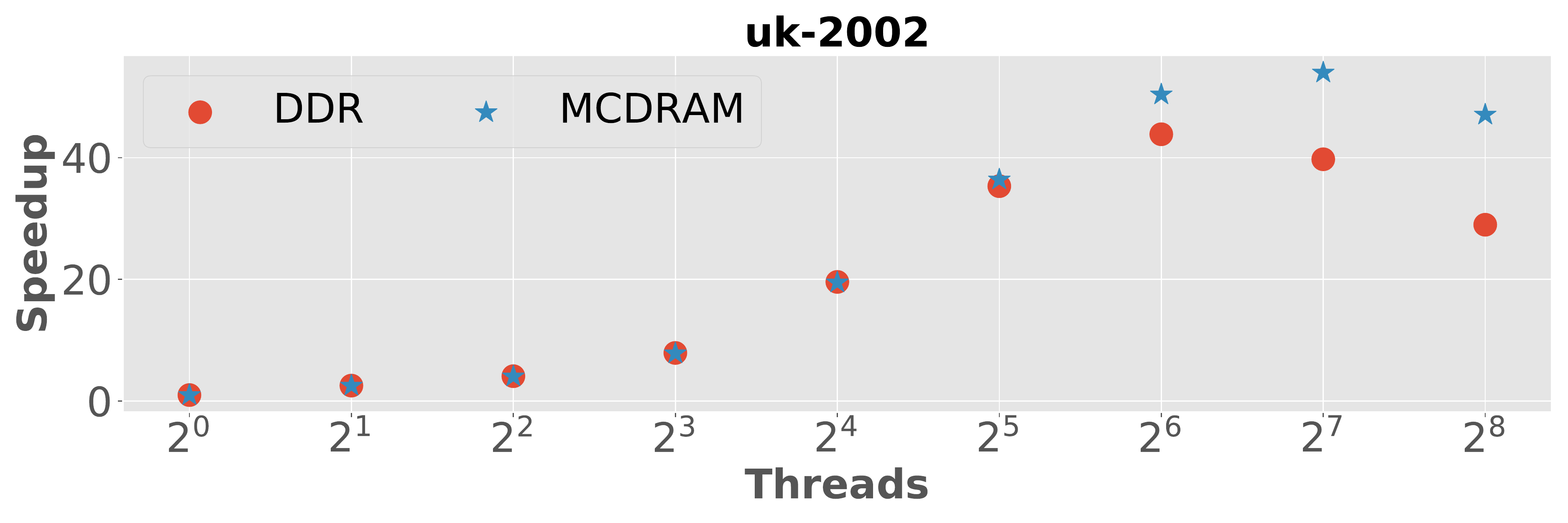}
 \includegraphics[width=0.66\columnwidth]{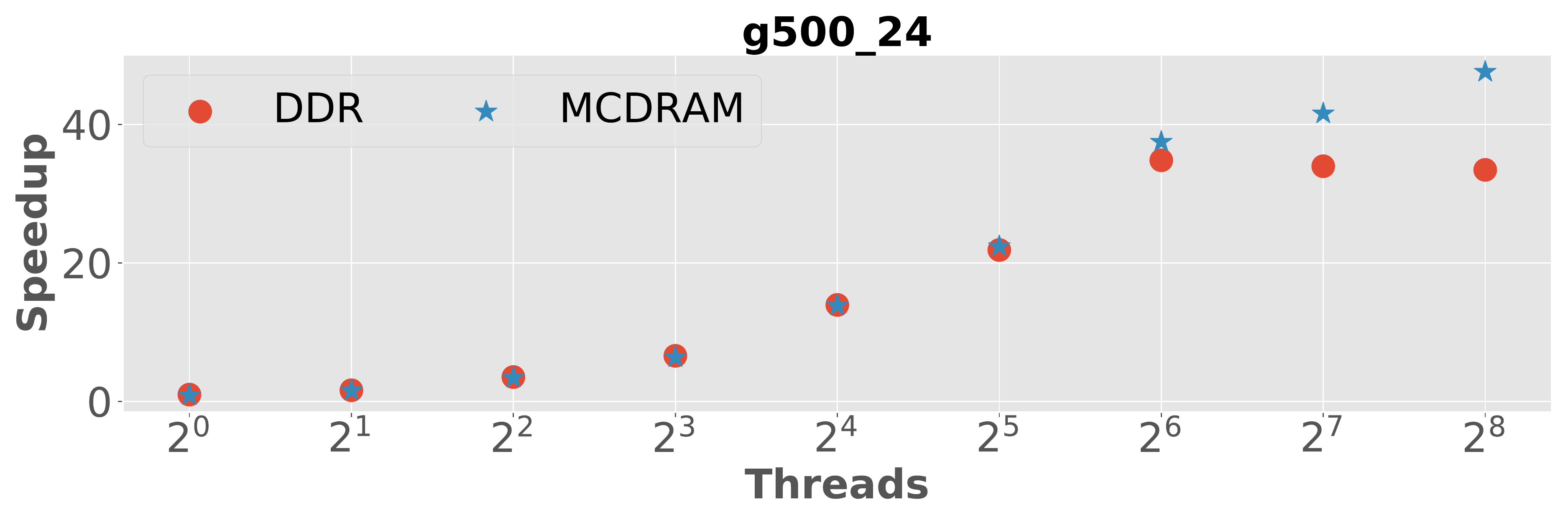}
}

\subfloat[\BC]{
 \includegraphics[width=0.66\columnwidth]{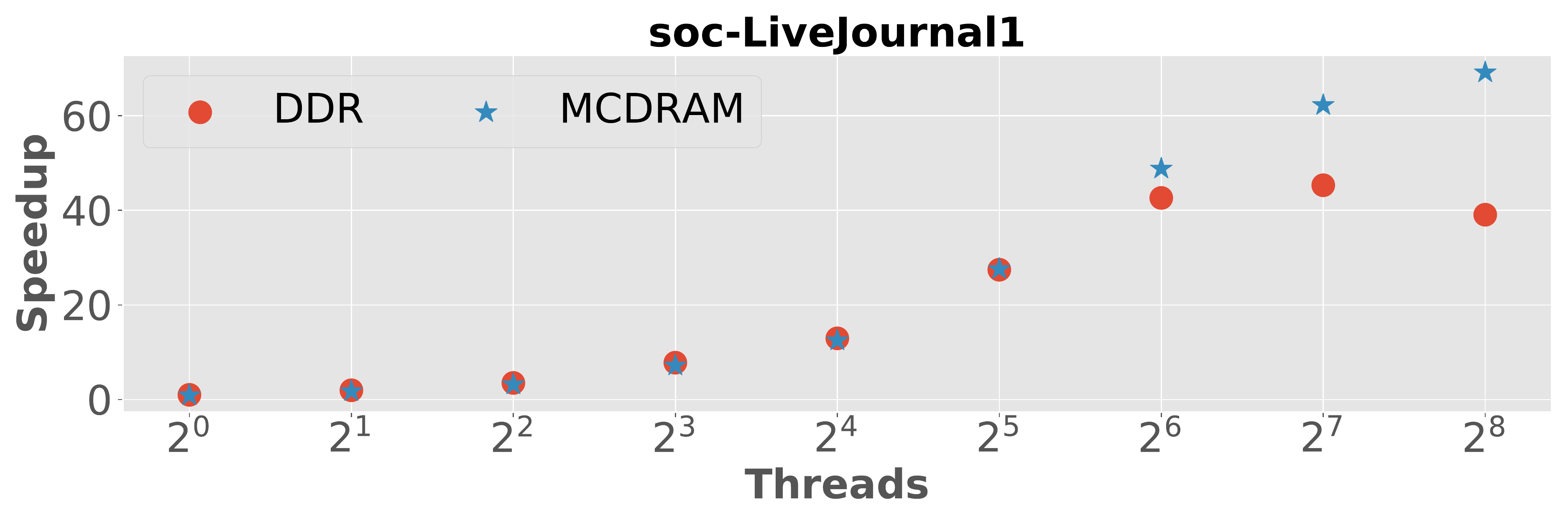}
 \includegraphics[width=0.66\columnwidth]{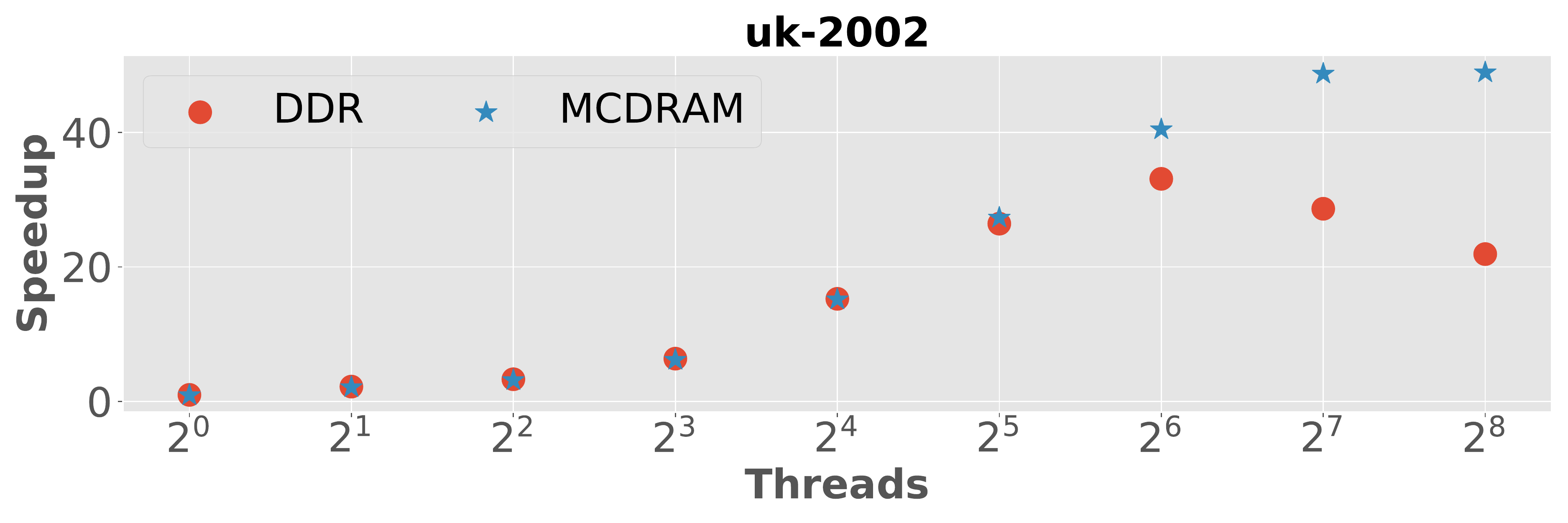}
 \includegraphics[width=0.66\columnwidth]{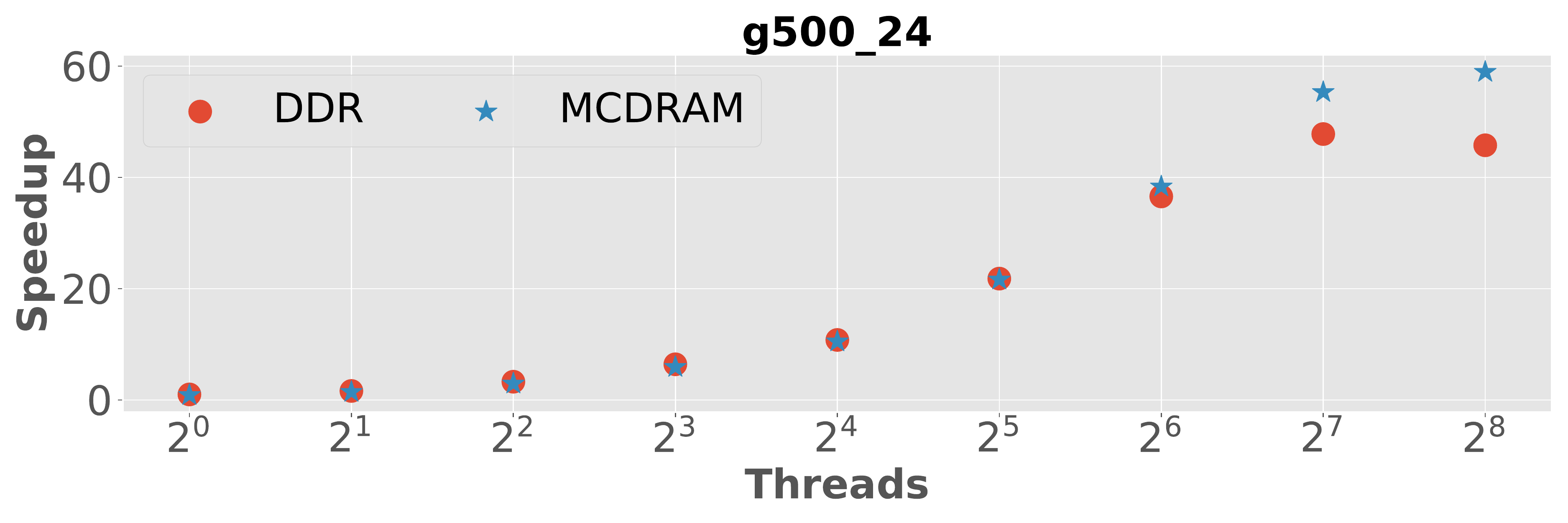}
 }

\subfloat[\PR]{
 \includegraphics[width=0.66\columnwidth]{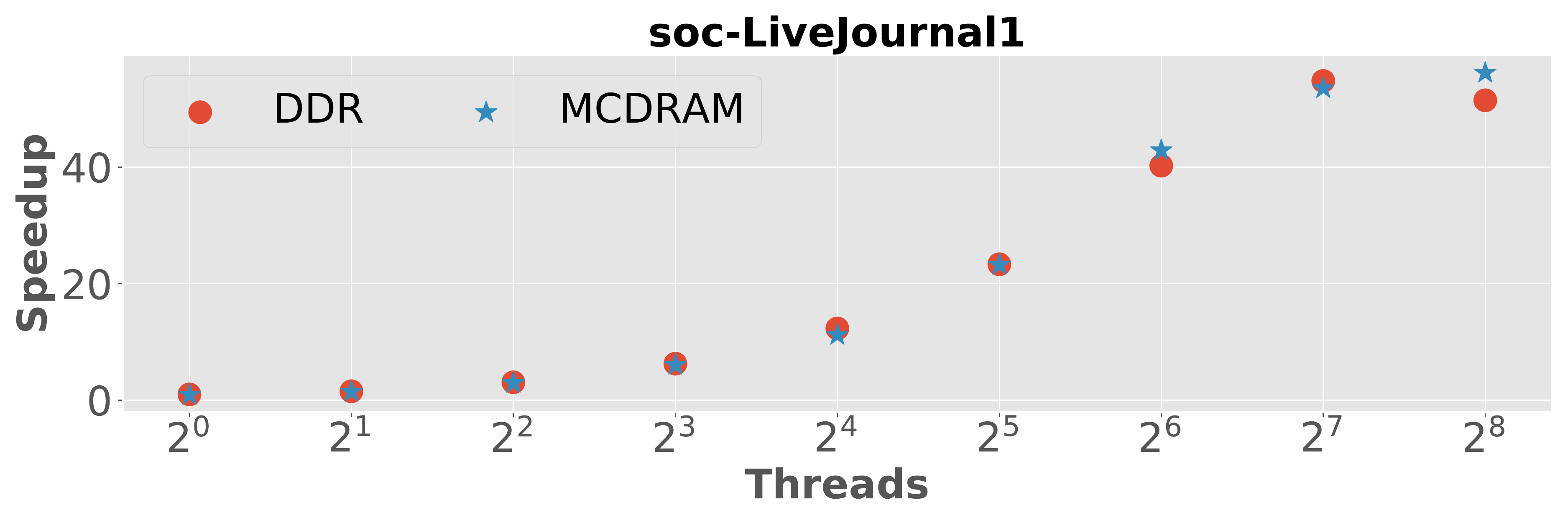}
 \includegraphics[width=0.66\columnwidth]{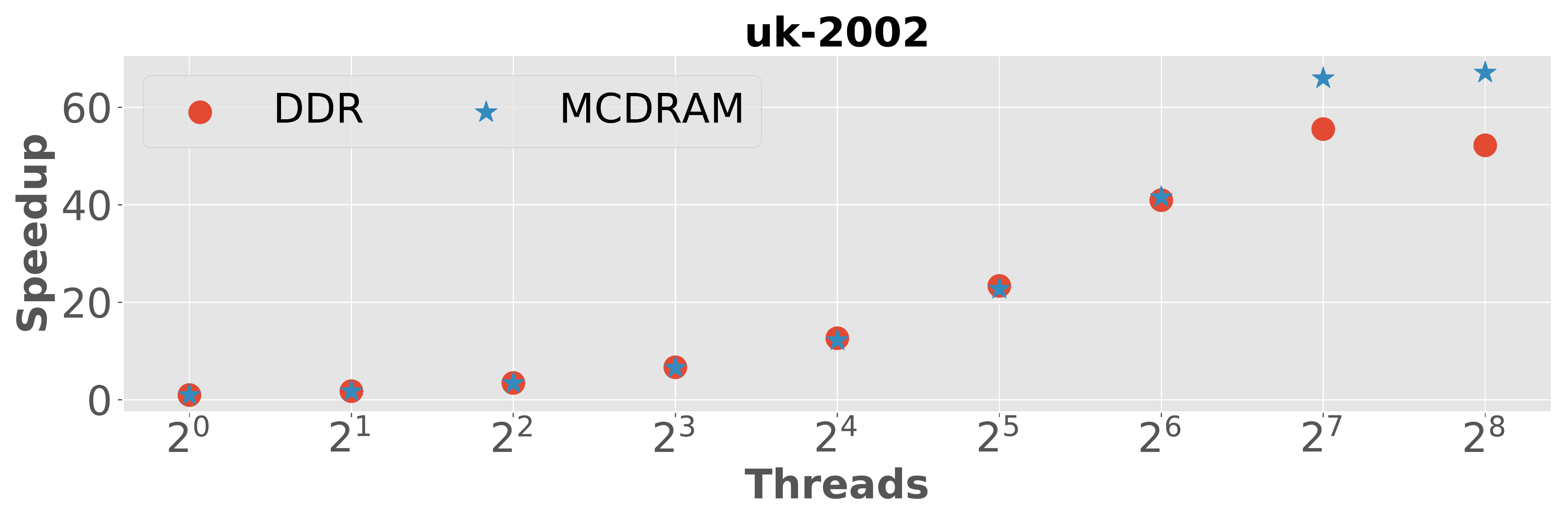}
 \includegraphics[width=0.66\columnwidth]{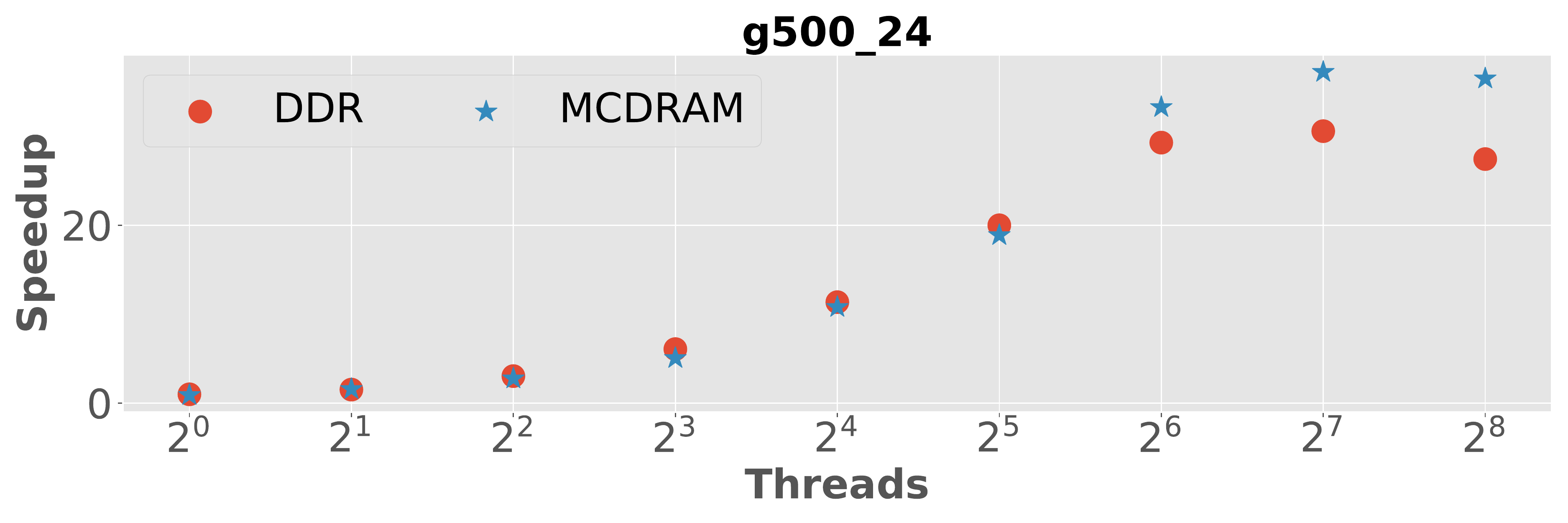}
 }

  \caption{Speedup relative to single-threaded execution on DRAM memory for GAP-BS \cite{shun2013ligra}. Execution times obtained on KNL using either DDR4 or MCDRAM.}

\label{fig:knl-gapbs}	

\end{figure*}

\begin{figure*}[t]
\centering
\subfloat[\BC]{
 \includegraphics[width=0.99\columnwidth]{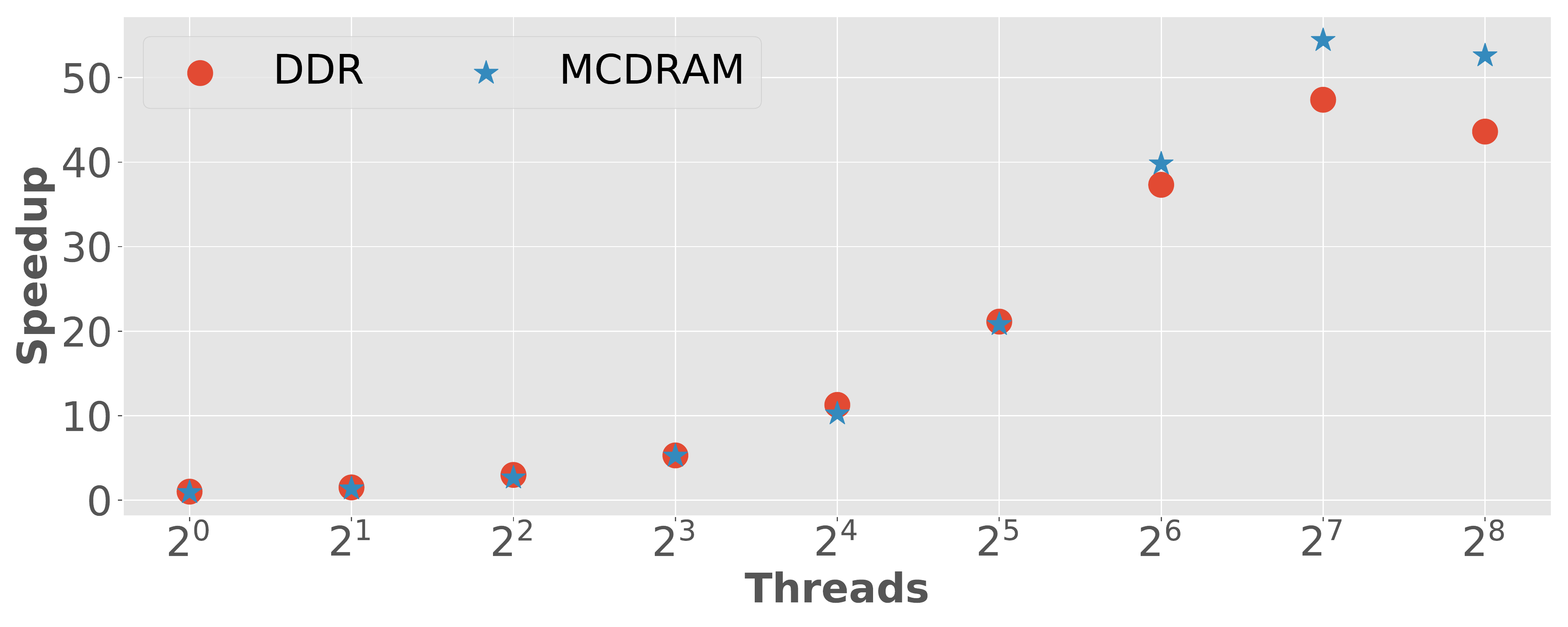}}
\subfloat[\PR]{
  \includegraphics[width=0.99\columnwidth]{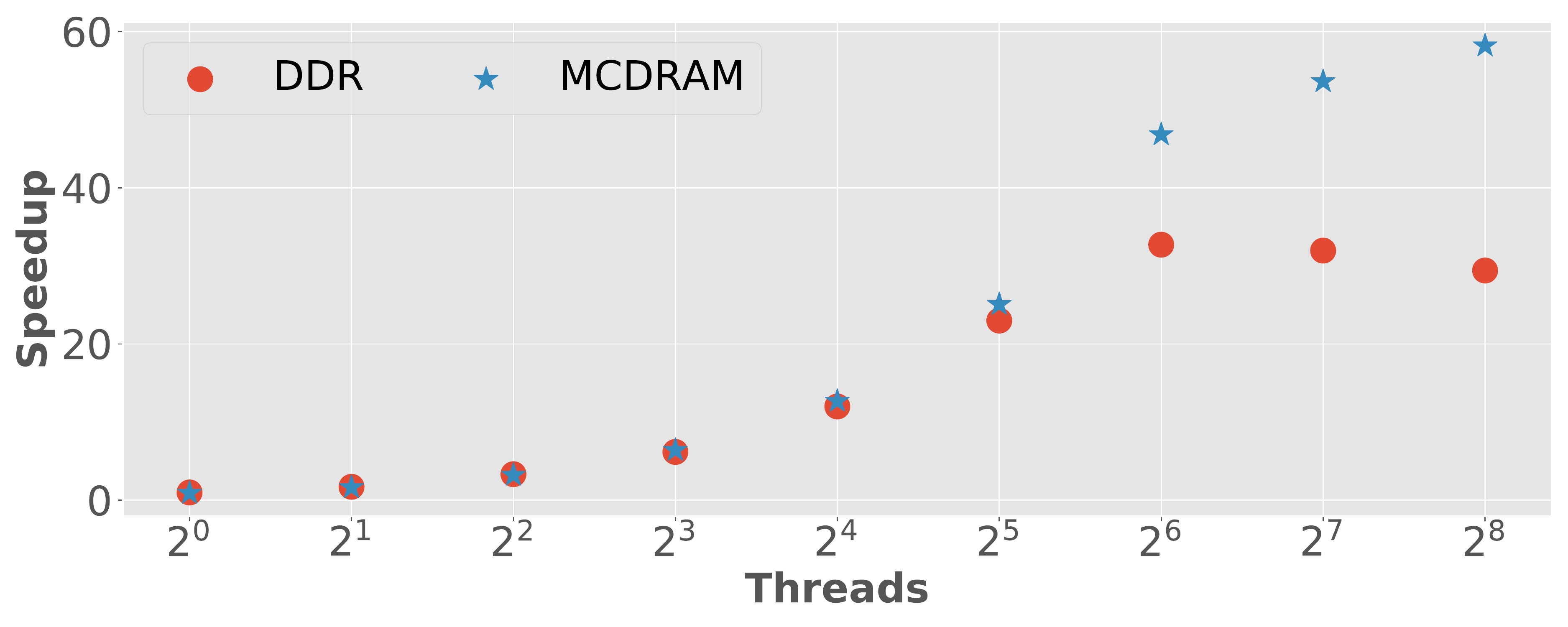}}

  \caption{Speedup relative to KNL single-threaded execution on DRAM memory for Ligra \cite{shun2013ligra} with either DDR4 or MCDRAM. The input graph is \textit{livejournal}.}

\label{fig:knl-ligra}	

\end{figure*}

\paragraph*{\bf Analysis}
Fig. \ref{fig:knl-gapbs} and Fig. \ref{fig:knl-ligra} depict speedups obtained from scaling up the number of threads for GAP-BS and Ligra, respectively. For these experiments, the exact same hardware, software, and input graphs are used with either DRAM (DDR4) or MCDRAM. 
The abscissa represents the number of threads and the ordinate is the speedup in comparison to sequential execution time with DDR4. 
We primarily report performance results for GAP-BS as it has faster execution times than Ligra. The faster execution times of GAP-BS allow us to further focus on the impacts of the memory subsystem rather on the load-balancing and runtime execution models of these two very distinct software frameworks. For GAP-BS we report on a wider range of input graphs, whereas for Ligra we report for the \textit{livejournal} graph.

Notice that for both frameworks, for both types of memory, and for most of the applications, the performance up to 32 threads is nearly identical. This is in part due to both memory subsystems having very similar latencies. As KNL has 4 threads per core, it is not atypical to see the performance drop when the number of threads is increased beyond the number of physical cores. 
For \BC there is little difference between the MCDRAM and DDR4 memory because BFS-like algorithms contain low-work phases that generate fewer memory requests per second. This phenomenon was also seen for \BFS in benchmarking by Peng et. al. \cite{peng2017exploring}. In contrast, for \PR the difference between the two memory technologies becomes apparent at 64 threads, and the performance gap is almost 2x for 256 threads. We do not show performance results for triangle counting since it generally exhibits both good spatial and temporal locality on current platforms.

\section{Results - GPU}
\label{sec:results-gpu}
\begin{figure}[tbp]
\centering
\subfloat[PageRank]{\includegraphics[width=0.99\columnwidth]{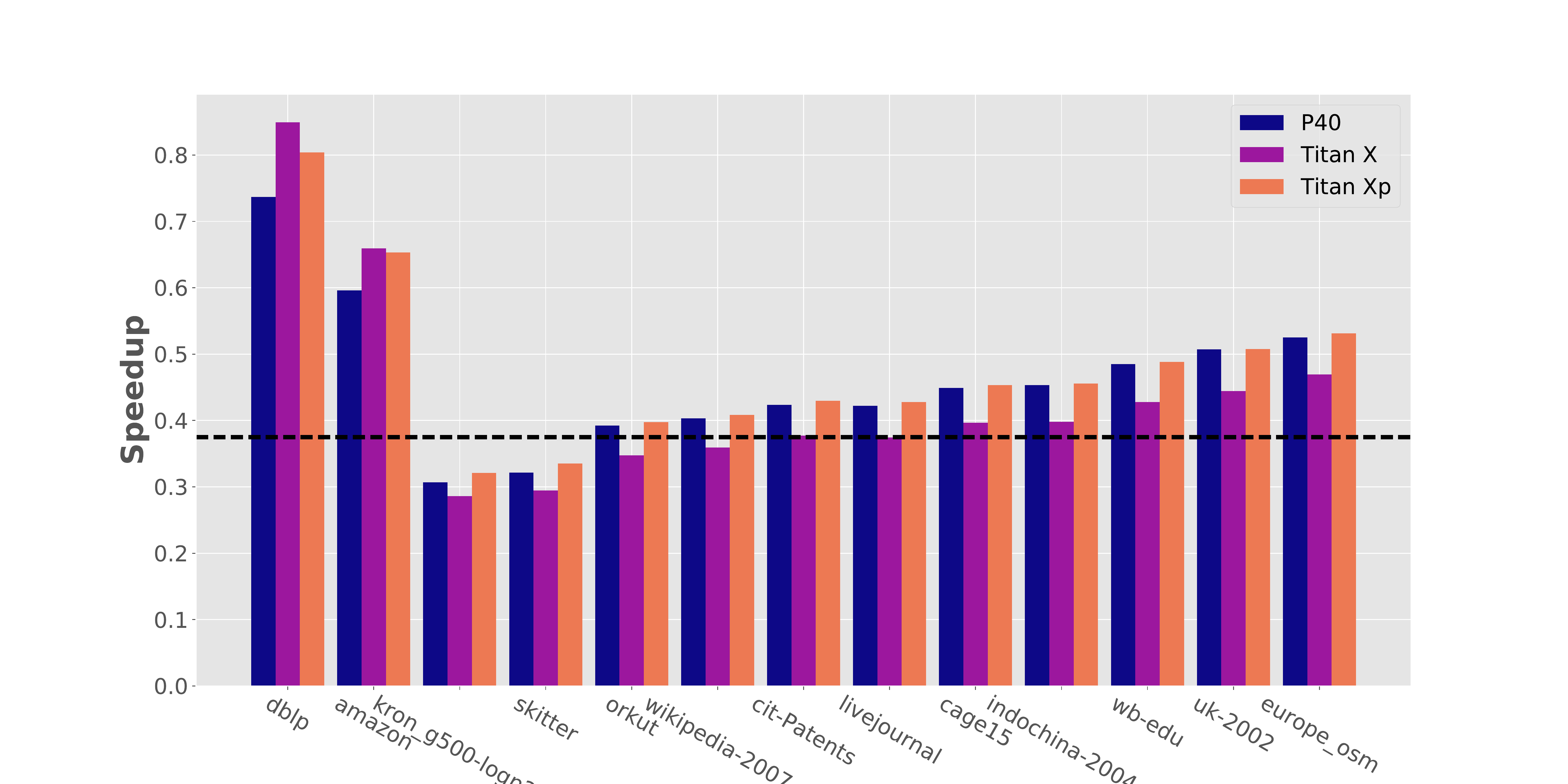}}

\subfloat[Betweenness Centrality]{ \includegraphics[width=0.99\columnwidth]{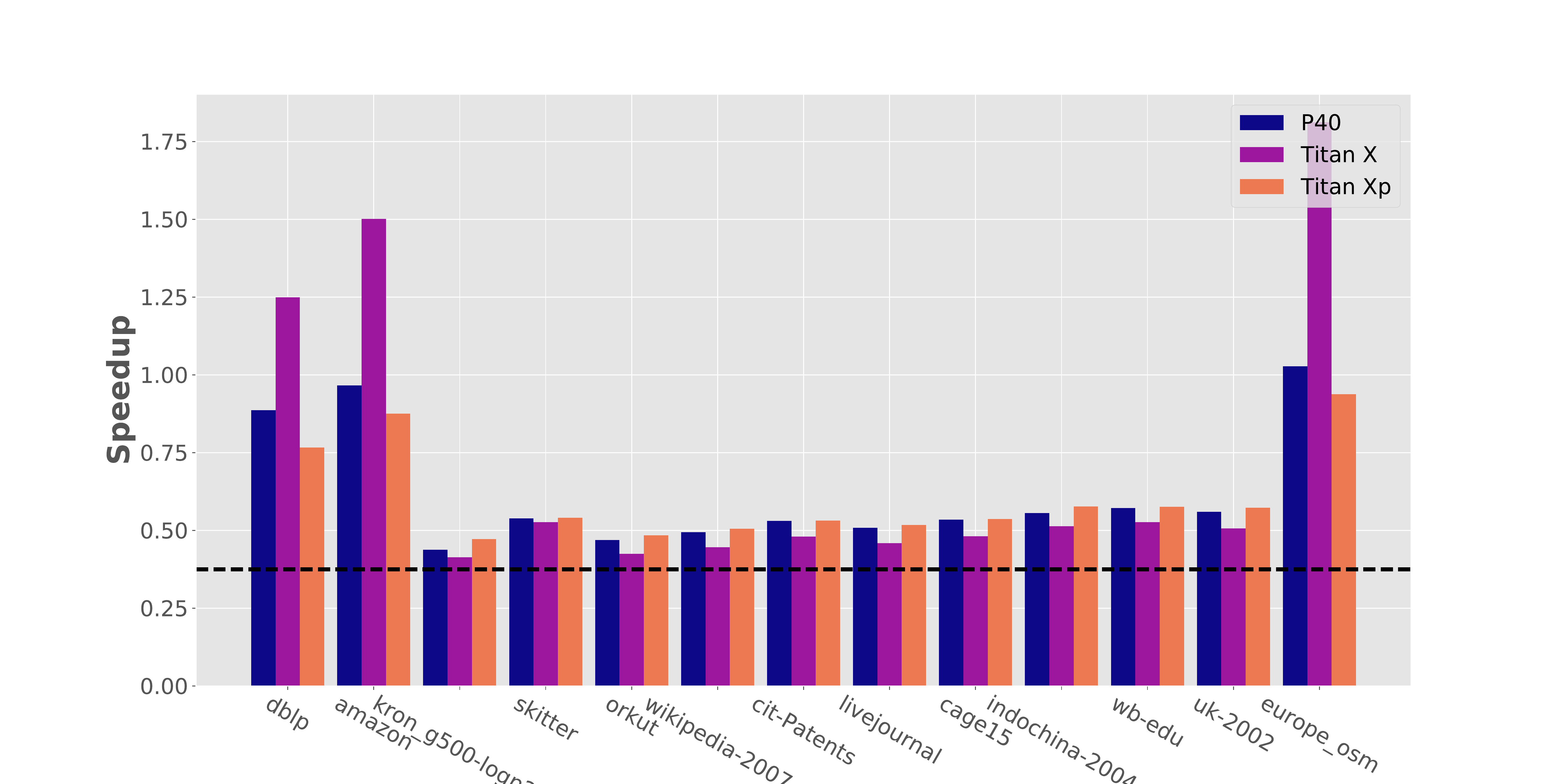}}

\subfloat[Shortest Path]{ \includegraphics[width=0.99\columnwidth]{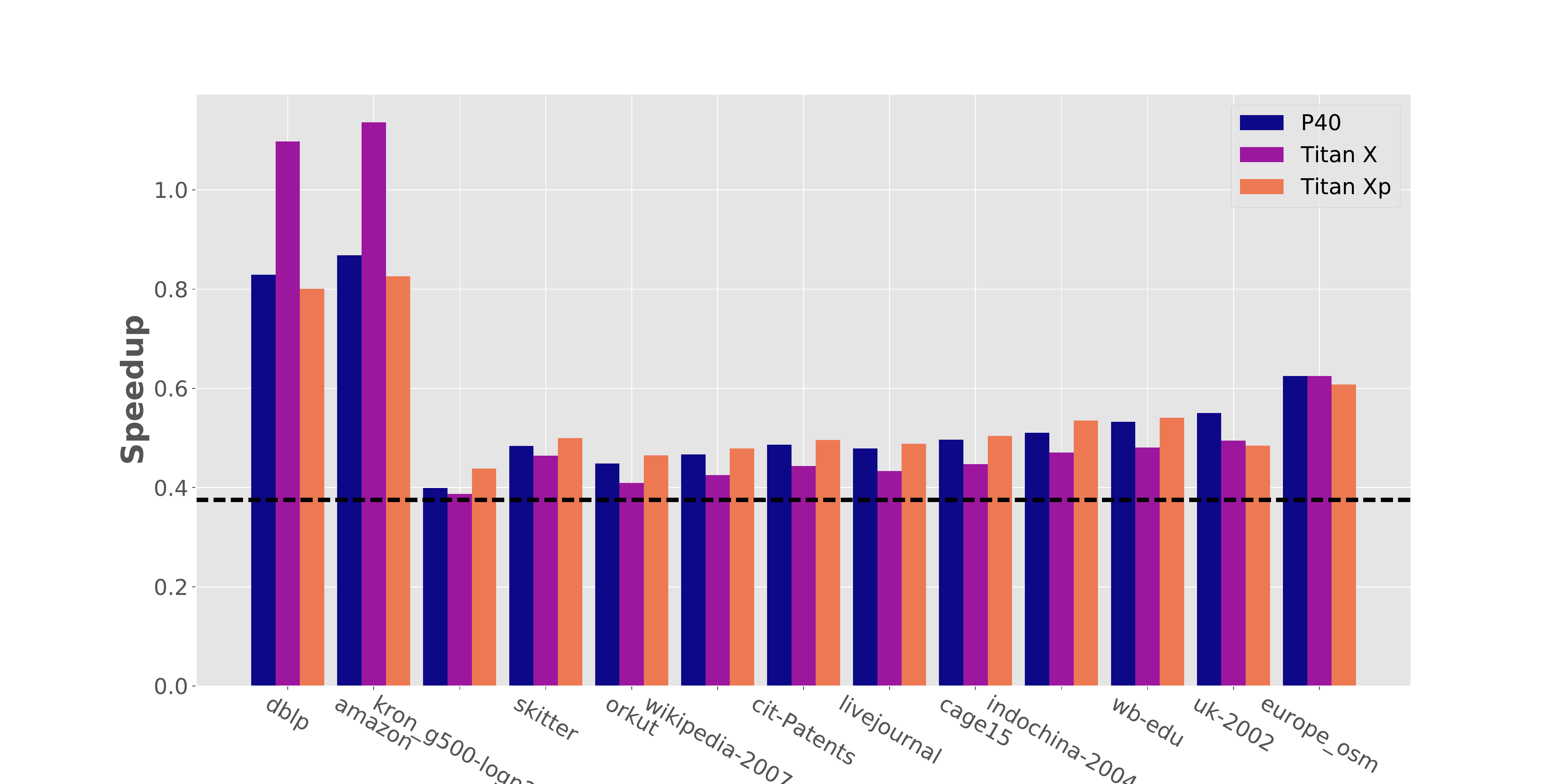}}

  \caption{Speedup using Gunrock \cite{wang2017gunrock} in comparison to a P100 GPU. Execution times are normalized to the P100's time (using HBM2). The dotted line $y=.375$ denotes the ratio of the number of memory channels for the listed GPUs with respect to the P100.} 

\label{fig:gunrock}	
\end{figure}

\begin{figure*}[t]

 \centering
 \subfloat[\PR]{
  \includegraphics[width=0.99\columnwidth]{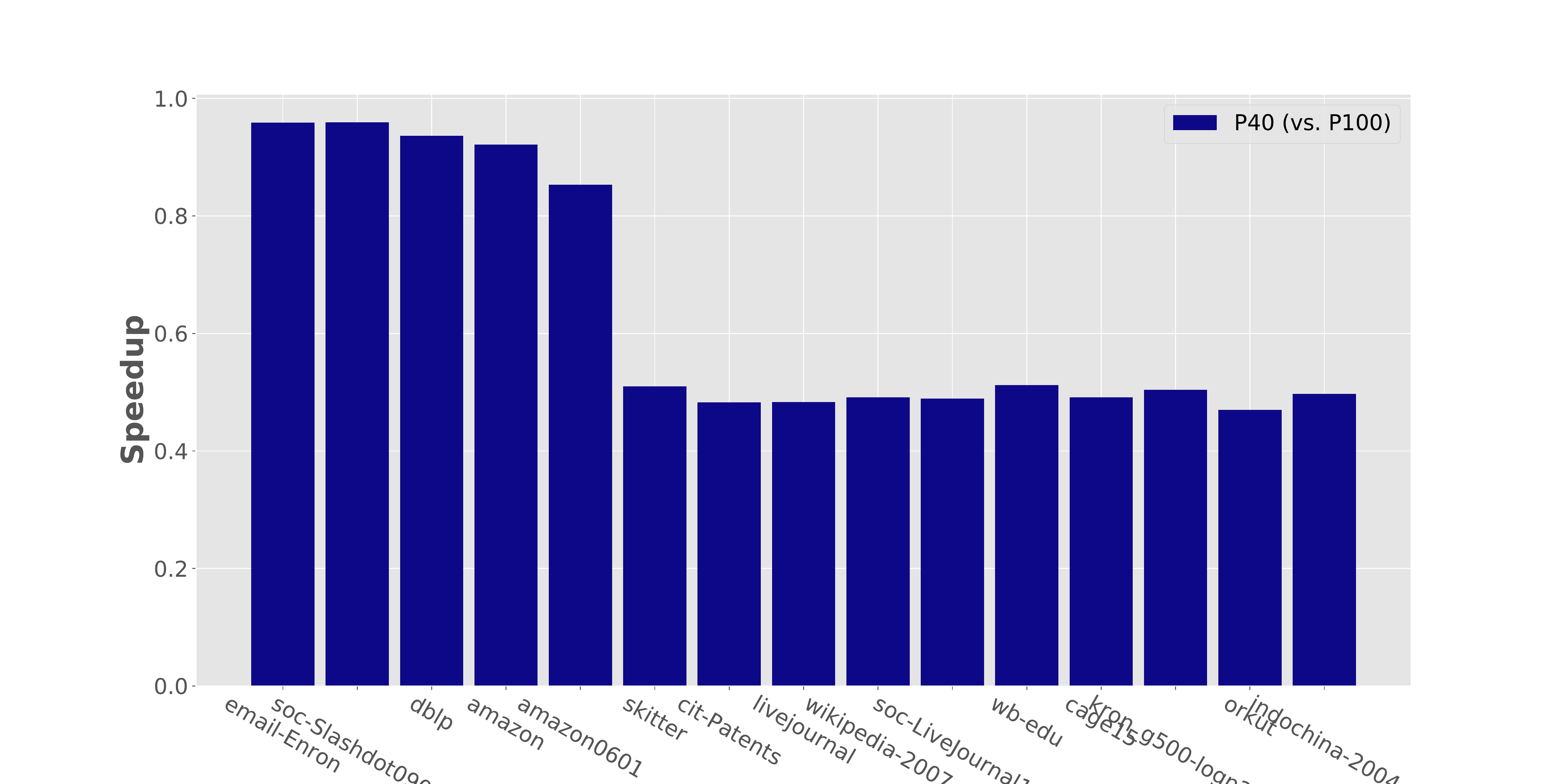}}
 \subfloat[\SSSP]{
   \includegraphics[width=0.99\columnwidth]{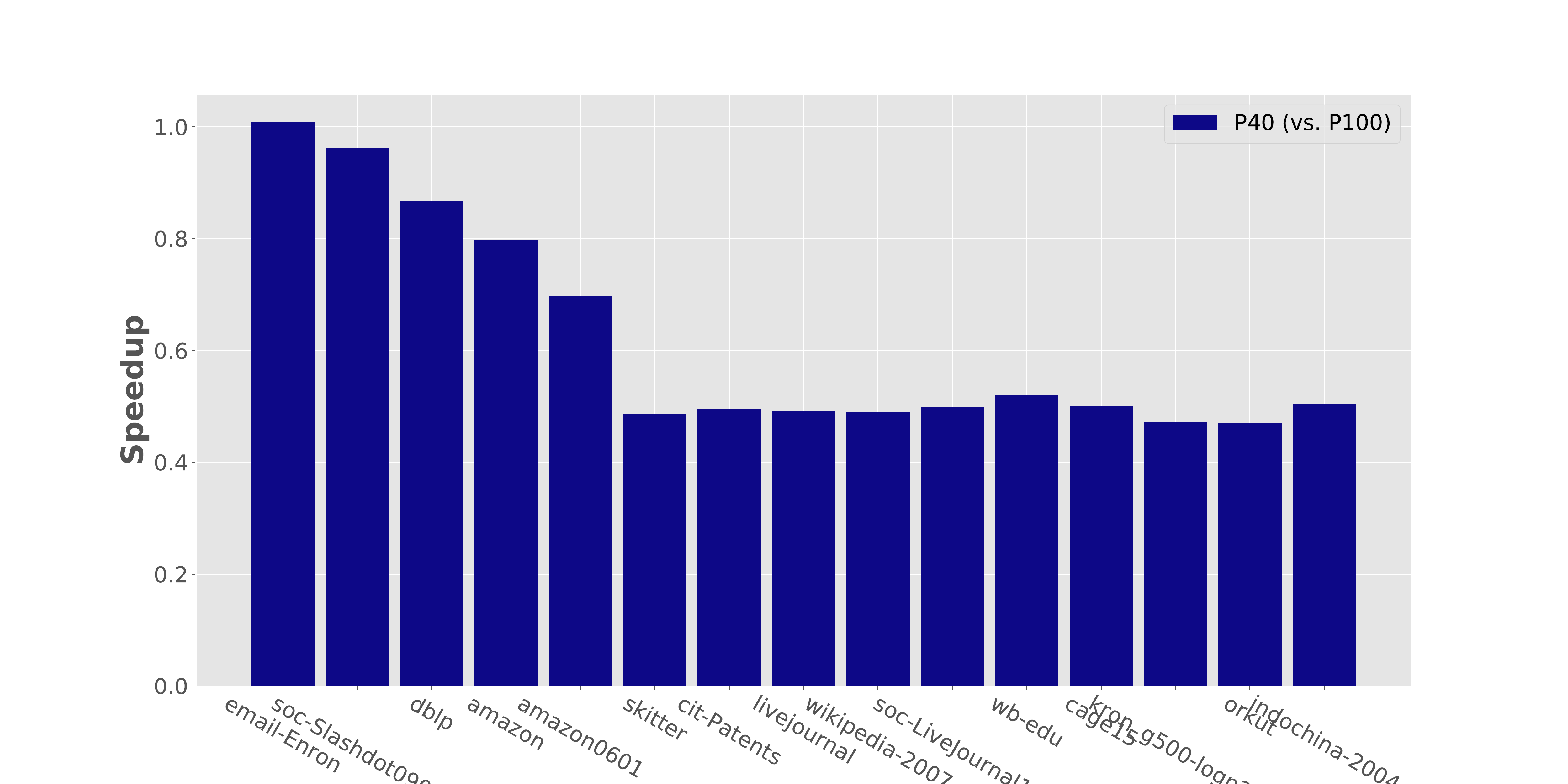}}

   \caption{Speedup for \PR and \SSSP using \NVG framework for the GPU.  Execution times are normalized to the execution time of the P100 GPU.}
 \label{fig:nvgraph}	
 
 \end{figure*}

\paragraph*{\bf System Configuration}

The specifications of the GPUs used in our experiments can be found in Table \ref{tab:gpu-cpu-systems}. The data transfer time between the GPU and GPU is not factored into runtime measurements, so we do not include the characteristics of the host systems and their CPU configurations. Experiments are primarily conducted on four different NVIDIA Pascal (same micro-architecture) GPUs. While the GPUs differ slightly in terms of core counts and clock frequencies (see Table \ref{tab:gpu-cpu-systems}), these devices have roughly equivalent FLOP rates. Benchmarking is also run on a NVIDIA V100 GPU (Volta micro-architecture) with HBM2 memory.

The Pascal-based GPUs have slightly different characteristics in terms of core counts, frequency, cache sizes, number of memory transactions per second, bandwidth, and {\bf memory channels}.  
Specifically, these differences include the following: 1) Core counts vary by $7\%$, 2) clock speed varies by $26\%$, 3) the P100 also has a larger L2 cache size (4 MB vs 3 MB), and 4) maximum memory bandwidth, which varies by more than $2\times$. Cache size is not as impactful for the larger graph inputs as these are several times larger than any current GPU's L2 cache. Thus, we can guarantee that a majority of the memory accesses will need to access global memory. 


The memory subsystems in the GPUs serves as the biggest distinction, and this varies between 345 GB/s on the P40 to the highest bandwidth of 720 GB/s on the P100. 
The NVIDIA P100 and V100 use High Bandwidth Memory (HBM2) and both have 32 memory channels serving 4096-bit total bus width. The remaining GPUS use GDDR5/GDDR5X memory and have 12 channels serving a 384-bit bus width. 
The number of memory channels for these processors is determined to the best of our ability using a recent NVIDIA presentation of high-bandwidth memory \cite{o2014highlights}. We also consulted per-channel width information reported in Micron Technology's GDDR5 and GDDR5X technical notes.

\subsection{Gunrock Analysis on Pascal and Volta GPUs}
We use the highly optimized Gunrock \cite{wang2017gunrock} framework to demonstrate that there is almost no difference for the three GDDR5 based GPUs despite there being up to a nearly $60\%$ difference in bandwidth. This analysis and comparisons to the P100 and V100 GPUs are used to show that these common analytics kernels are not primarily \textit{bandwidth-bound}.

\paragraph*{\bf Experiment setup}
To measure the impact of the memory channels on the performance of graphs on the GPUs we use both Gunrock, and we also report results for \NVG, a linear-algebra based graph analytics library by NVIDIA.

For Gunrock, we measure the execution time (excluding data transfer) for the following analytics: \BC, \SSSP, and \PR. Runtime results for \BC and \SSSP are averaged for the 200 top-degree vertices in the graph. Results for \PR are averaged for 5 runs of each kernel, where each run has 50 fixed iterations.
Using \NVG we measure the execution time for \SSSP and \PR using similar settings. 

\paragraph*{\bf Gunrock Analysis}
Fig. \ref{fig:gunrock} depicts the performance of several graph algorithms using Gunrock. 
The abscissa represents the graph used in the experiment.
The ordinate represents the normalized speedup execution with respect to the NVIDIA P100 PCI-E based GPU, the only NVIDIA Pascal GPU to have high-bandwidth memory. 

For most algorithms and inputs, the P100 GPU outperforms the remaining GPUs by a significant factor - despite the other GPUs supporting a larger number of memory transactions per second. There are a few instances where the GDDR5-based GPUs outperform the P100. In most cases, this occurs for the smaller graphs where the data structure used by the analytics can mostly fit into the GPU's LLC. There are a very few  cases where the larger graphs perform better on the GDDR5-based GPUs. As these experiments are not optimized to take advantage of a specific GPU, the difference in execution might be associated with load-balancing. These few cases require additional investigation.

For three of the benchmarks (\PR, \BC, and \SSSP), the P100 outperforms the other benchmarks by about 2x-3x. Recall that the P100 has 32 memory channels whereas the other GPUs have 12. We have added a line at $y=\frac{12}{32}=0.375$ (in all the subplots of Fig. \ref{fig:gunrock}) to indicate this ratio. Note that the bars of the GDDR5-based GPUs are fairly close to this curve. This is especially surprising given the fact that the Titan XP has almost $60\%$ more bandwidth than the P40, while the difference in their execution time is clearly not $60\%$. Further, the Titan XP has $75\%$ of the peak bandwidth fort the P100. If these graph algorithms were solely bandwidth-bound, then we could expect the execution times to be correlated. Rather, it seems that, across all the GPUs, the ability to handle saturation of memory subsystems via additional channels is the dominant factor in the performance of the algorithms.

\paragraph*{\bf NVIDIA (VOLTA) V100 GPU} In the above analysis we primarily focus on the Pascal based GPUs as these share a lot of common traits. Experiments with the V100 GPU (not shown) show that the performance of the V100 is quite similar to the P100 GPU despite having an additional $30\%$ cores and extra $25\%$ bandwidth. This evaluation demonstrates that performance is constrained by the number of memory channels rather than by other factors.

\paragraph*{\bf nvGraph Analysis}

We also run experiments using \NVG \cite{nvGraph}, specifically for \PR and \SSSP. NVGraph implements graph algorithms using linear algebra based operations (as discussed in the GraphBLAS standard \cite{kepner2016mathematical}). 
Similar trends are seen as with the results for P40 and P100 GPUs in Fig. \ref{fig:nvgraph}, which shows up to a 40\% performance penalty for the P40 GPU when compared to a P100 baseline. 
By using nvGraph, in addition to Gunrock, we show that the performance impacts of many small, irregular accesses is not dependent on the programming and load-balancing scheme that is use in Gunrock, and we demonstrate that this performance penalty for high-bandwidth, limited memory channel systems also can be found in sparse matrix-based implementations of graph analytics.

\begin{table}[t]
\centering
\caption{GPU Bandwidth Analysis With Synthetic Micro-benchmarks.}
\scriptsize

\begin{tabular}{|c|c|c|c|c|} \hline
& P100 & P40 & Titan Xp & Titan X  \\ \hline
Peak bandwidth (GB/s) & 720 & 345 & 547.7 & 480  \\ \hline
Coalesced read (GB/s) \cite{danalis2010scalable} & 573 & 230 & 316 & 270 \\ \hline
Coalesced write (GB/s) \cite{danalis2010scalable} & 432 & 249 & 312 & 266 \\ \hline
Random read (GB/s) & 17.1 & 9 & 9 & 7.7 \\ \hline
Random write (GB/s) & 7.4 & 3.9 & 8.4 & 7.6 \\ \hline

\end{tabular}
\label{tab:streaming}
\end{table}

\subsection{Synthetic Benchmark Analysis}

Table \ref{tab:streaming} depicts the performance of the GPUs running two synthetic benchmarks: 1) the SHOC benchmark \cite{danalis2010scalable}, which tests the GPUs bandwidth capabilities for regular memory access patterns and 2) random memory access benchmarks that measure random reads and writes in an array. 
The latter is our own implementation, which performs reads and writes to addresses determined by an efficient FNV-1 \cite{noll1991fnv} hash. Bandwidth reported for random reads and writes is \textit{effective bandwidth}, or simply the total number of memory accesses (scaled by data type size) divided by execution time. 
The coalesced read/write benchmarks show relative performance consistent with peak bandwidth differences across the different GPUs. However, our benchmarks suggest that the P100 is roughly twice as effective as other GPUs on random reads. For random writes, the P100 achieves comparable bandwidth to the Titan Xp and Titan X, a ratio that is $2\times$ more than the P40.

\section{PageRank - In-depth Case Study}
\label{sec:model}

The experiments in previous sections look at a cross-section of behavior for several widely used graph algorithms. In this section, we take an in-depth look at separating the computation- and communication-related behaviors of one specific graph algorithm, \PR, and we attempt to distinguish between random and consecutive memory accesses for sparse problems. Random memory accesses are typically bounded by the memory subsystem while consecutive accesses tend to be compute bound due to cache-related data reuse. We further evaluate \PR and speculate on how its performance might be changed by more numerous but narrower memory channels using a simple performance model and \textit{gap analysis} experiments.

\begin{algorithm}[t]

\tiny

    \For {$v \in V$}
    {
        $PR_{curr}[v] \leftarrow 0$;  
        $C[v] \leftarrow Pr_{prev}[v]/deg[v]$;
    }

    \For {$v \in V$}
    {
        // ``Streaming edge'' access //
        \For {$v \in Neighbors[u]$} 
        {
        	$PR_{curr}[v] \leftarrow PR_{curr}[v] + C[u]; $ // Random access \\

        }
    }
    \For {$v \in V$}
    {
        $PR_{curr}[v] \leftarrow \frac{1-d}{NV} + d\cdot PR_{curr}[v];$ 
        $PR_{prev}[v] \leftarrow Pr_{curr};$
    }

\caption{Page-Rank pseudo code.}
\label{alg:pr}

\end{algorithm}

\begin{figure}[t]
\centering
  \includegraphics[width=0.85\columnwidth]{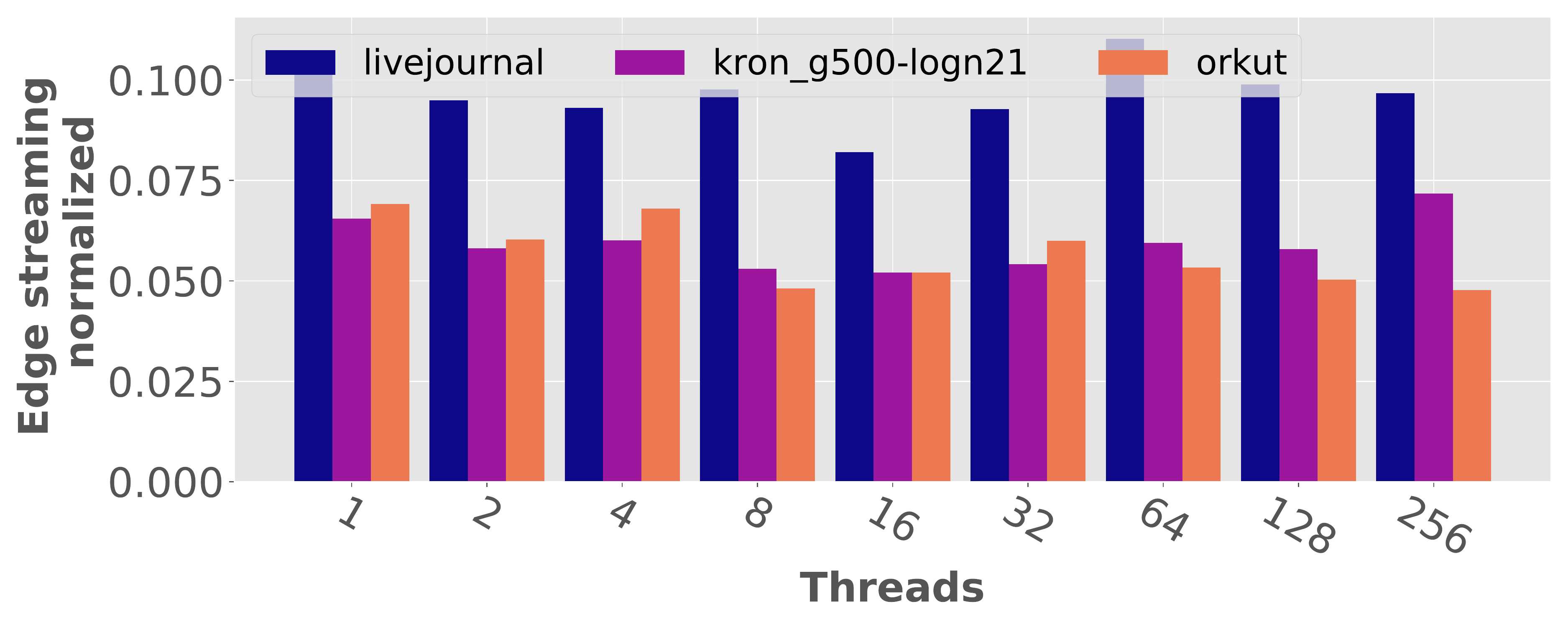}

  \caption{Adjacency traversal-only execution time as a ratio of PageRank execution time for DDR4 memory on KNL.}

\label{fig:streaming-pr} 
\end{figure}





\subsection{Experiment Setup}

We choose \PR as our in-depth case study, as it shows divergent MCDRAM and DRAM performance in our earlier evaluation (Section \ref{sec:results-cpu}). 
For the sake of simplicity, we follow a very simple \PR model as defined in Alg. \ref{alg:pr} for our analysis. Our reasons are two fold: 1) the two key types of memory accesses which we define below are also found in other optimized algorithms and 2) several of the optimized algorithms, such as \cite{zhou2017design,beamer2017reducing}, have been designed to reduce the number of random memory accesses, which makes it harder to stress and evaluate the memory system with this type of algorithm.

We use the PageRank implementation from Ligra but makes a few small changes to its implementation. Ligra's variation is entirely pull-based, which means that each vertex is responsible for updating its own PageRank value. For simplicity, we fix the number of PageRank iterations at 20. As part of this experiment, we updated the Ligra implementation to use $O(|V|)$ division operations instead of $O(|E|)$. This change removes computational overheads from our analysis and makes it easier to identify performance problems. 
Performance is evaluated using the $liveJournal$ graph, and tests are executed on the KNL platform described earlier in this paper. 


\subsection{Random Versus Consecutive Memory Accesses}

In this set of evaluations, we differentiate between consecutive or ``\textit{graph streaming}" accesses and random memory accesses. Consecutive accesses focus on fetching in elements of the graph, typically edges from an adjacency list as found in CSR (which is also used by Ligra). These have good spatial locality.
On the other hand, ``\textit{random memory}'' accesses are more data dependent accesses that are typically used by the analytic itself to update and store specific values. For \PR, iterating through adjacency lists forms the streaming-like memory accesses, while reading and updating vertex values produces sparse memory accesses. For PageRank, each ``\textit{graph streaming}" access roughly corresponds one-to-one with a random memory access as edges are fetched in the graph streaming phase and then used to update \PR values with random accesses. 

Fig. \ref{fig:streaming-pr} depicts the percentage of time spent streaming in the graph from the total execution time for KNL's DDR4 DRAM memory. In practice, this phase includes the entire \PR computation excluding the random memory accesses. While there is a slight difference between the percentage across the number of threads, the overall time spent in this phase is relatively small (less than 10\% for $liveJournal$ and two other input graphs) in comparison to the time spent in the random access part. This result demonstrates the benefits of prefetching and cache reuse for traversing the adjacency array. As part of this experiment, we verify that the bandwidth is saturated and see empirically that we get 70-80\% of peak DRAM bandwidth for high thread counts. We also verify that the number of cache misses is identical for various threads counts and the two memory subsystems (MCDRAM and DDR4). This is expected behavior as the cache misses are not architecture dependent but are input dependent. In part, this also shows that there is not much data reuse across the threads, meaning that there is low spatial and temporal locality.
When combined with the percentage of time spent in the ``\textit{graph streaming}" phase in Fig. \ref{fig:streaming-pr}, these experimental results indicate that random memory accesses interacting with the 
DRAM memory subsystem are the cause for performance drop-off seen for DDR4 results in Fig. \ref{fig:knl-gapbs} and Fig. \ref{fig:knl-ligra}.

\begin{figure*}[ht]
\centering
\subfloat[DRAM - Graph Streaming]{
 \includegraphics[width=0.99\columnwidth]{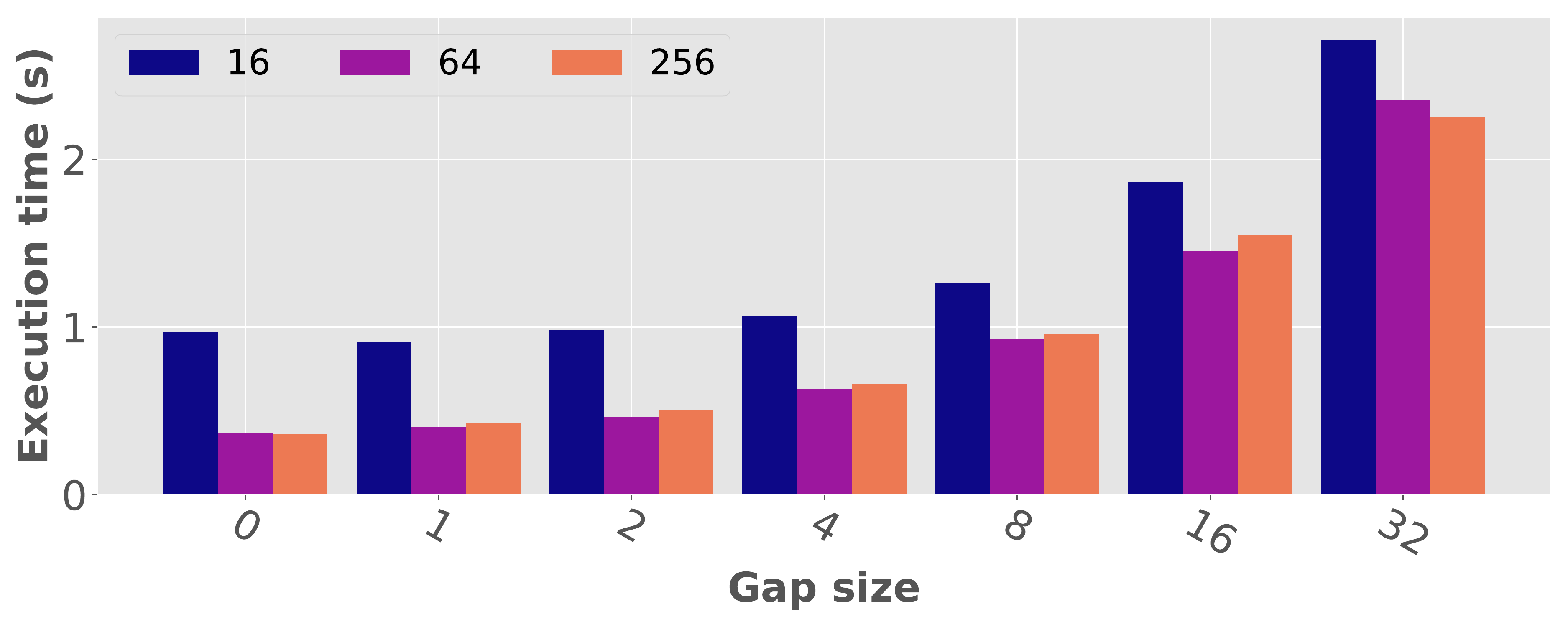}}
\subfloat[MCDRAM - Graph Streaming]{
 \includegraphics[width=0.99\columnwidth]{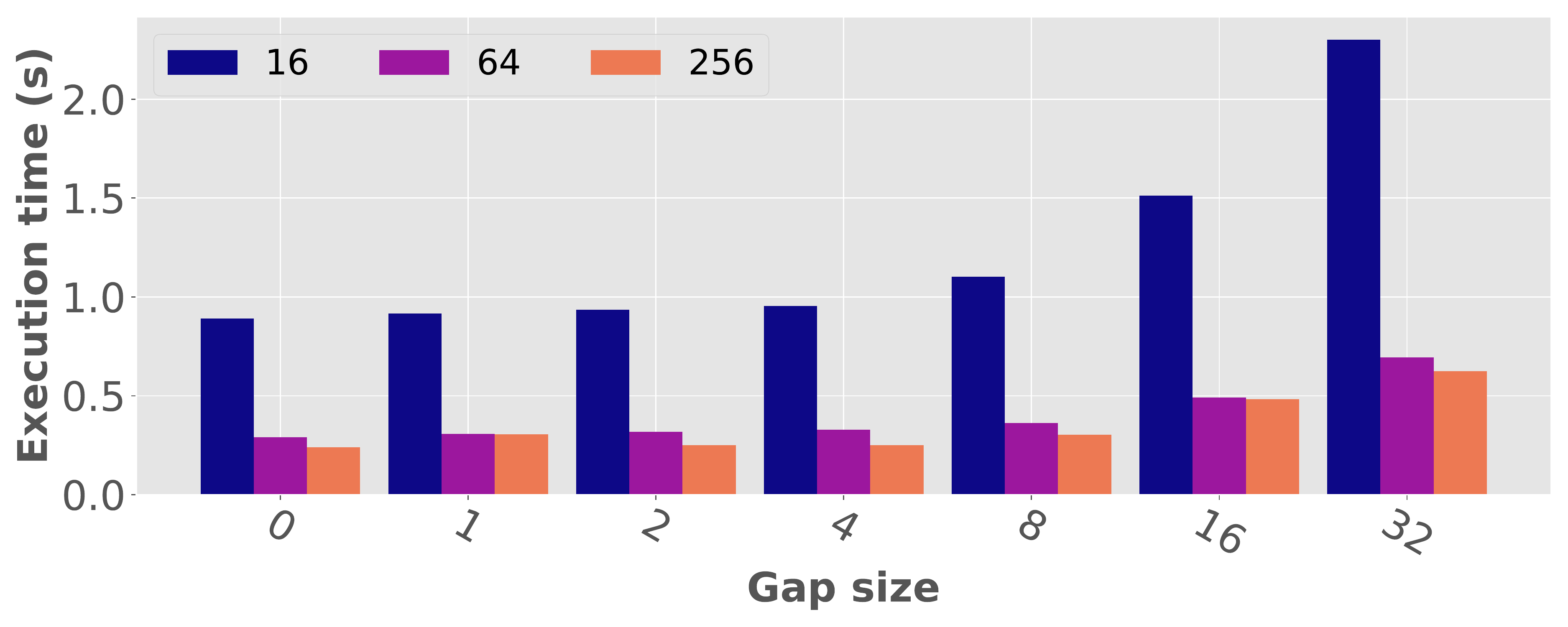}}

\subfloat[DRAM - Overall]{
 \includegraphics[width=0.99\columnwidth]{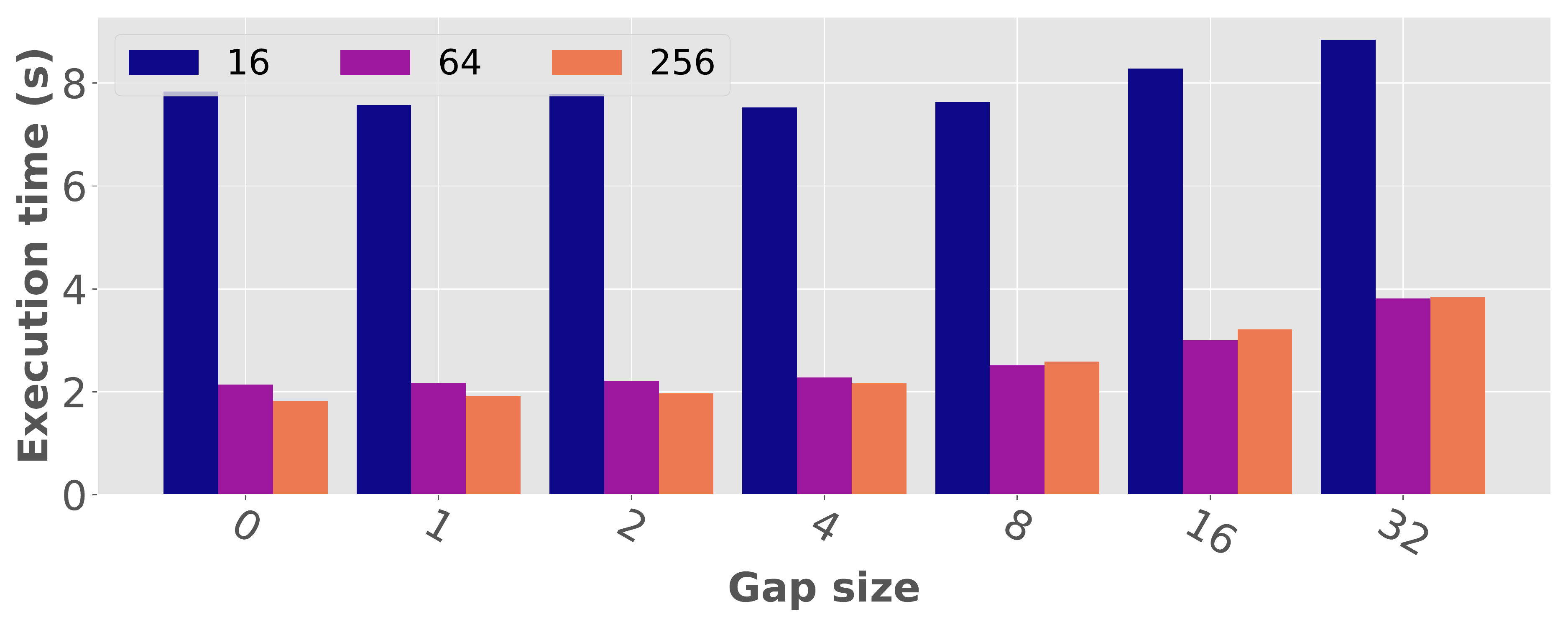}}
\subfloat[MCDRAM - Overall]{
 \includegraphics[width=0.99\columnwidth]{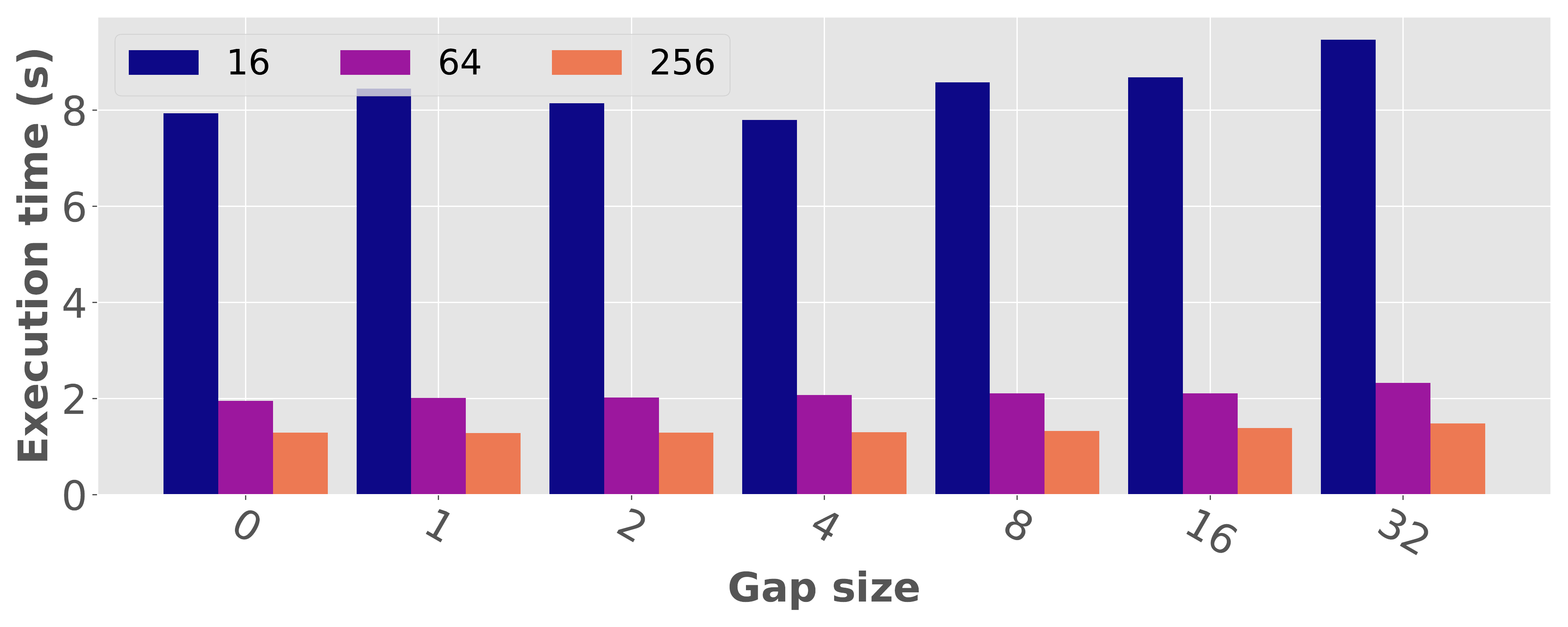}}

  \caption{Increasing gap size (powers of 2) vs. execution time for Ligra PageRank using either DRAM or MCDRAM memory on the KNL. Gap size of 16 onwards emulates cache line accesses with no data reuse. Top row shows times for graph streaming while the bottom row shows overall execution times.}

\label{fig:gap-time-pr} 
\vspace{-5mm}
\end{figure*}

\begin{figure*}[ht]
\centering
\subfloat[DRAM]{
 \includegraphics[width=0.99\columnwidth]{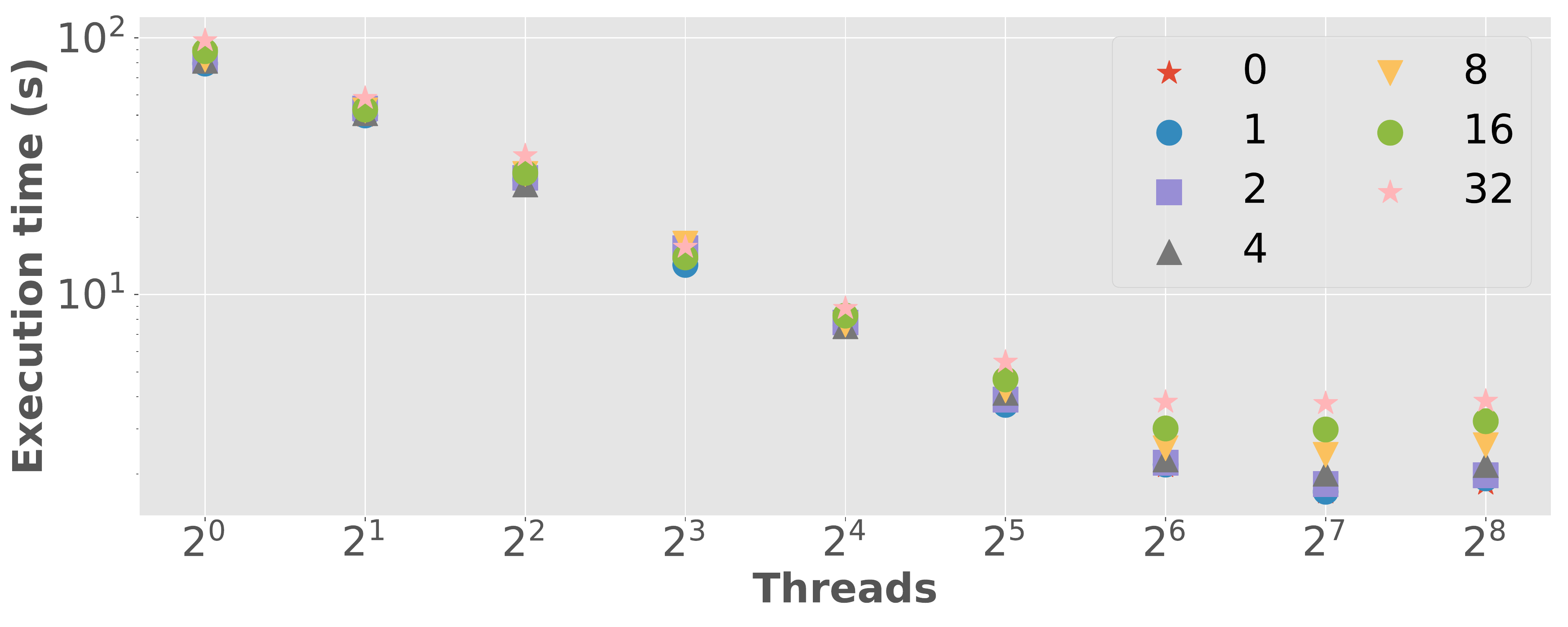}}
\subfloat[MCDRAM]{
 \includegraphics[width=0.99\columnwidth]{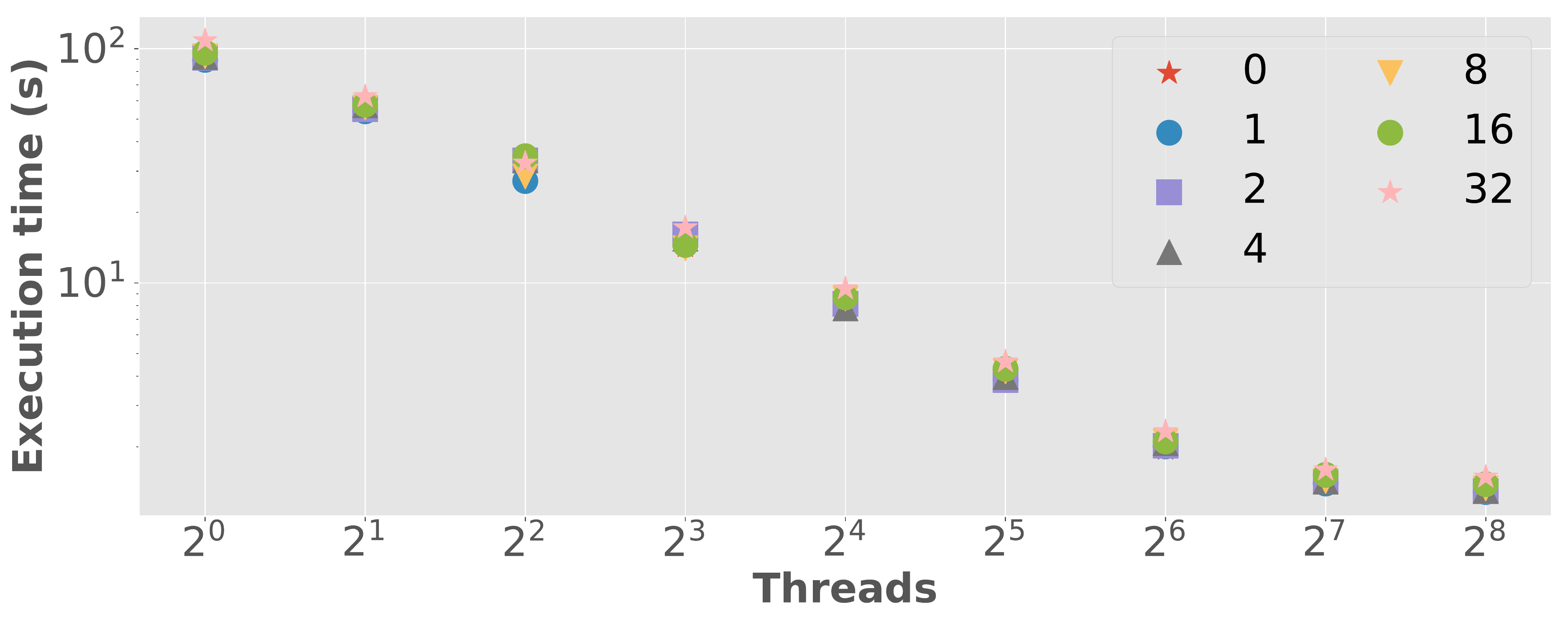}}

  \caption{Overall execution time (log-scale) as a function of number of threads used, for PageRank. The different curves represent performance scaling for different gaps introduced into the graph. The two plots compare using DRAM and MCDRAM memory for the same experiments.}

\label{fig:gap-scaling} 
\vspace{-5mm}
\end{figure*}

\subsection{Modeling the Addition of Narrow Channels}

As mentioned in Section \ref{sec:related}, several proposed technologies like Half-DRAM and Fine-grained DRAM have introduced the idea of additional memory channels with narrower widths per channel. 
In this section, we focus on understanding how these systems would behave by devising a simple performance model to help evaluate the performance behavior with narrow memory channels and fine-grained accesses. We use \PR as our baseline.

For many random memory accesses, a cache line is fetched in but only a single data element may be used. For example, a random memory access may bring in a whole cache line of 64 bytes but only use 4 bytes, which makes the utilization of that cache line $6.25\%$. As such, a narrower memory channel is not likely to negatively impact the performance for these accesses. In contrast, the graph streaming phase uses a contiguous adjacency array, and it is very likely that the entire cache line will be reused. Fig. \ref{fig:streaming-pr} indicates that a relatively small amount of time is used overall for adjacency arrays, so narrower channels are not likely to increase the overall time for the graph algorithm due to the dominance of the random access phase. 

To formalize these trade-offs in a simple model, let $T = R + S$ be the total time attributed to random accesses $R$ and streaming accesses $S$. We assume that introducing narrower channels increases streaming overhead by a fixed factor $K$, while enabling a factor of $C$ additional channels. Then the speedup for memory accesses from more narrow channels is given by: 
\begin{equation}
Speedup = \frac{1}{\frac{1}{C} \frac{R}{R+S} + \frac{K}{C} \frac{S}{R+S}}
\label{eq:perf}
\end{equation}

Intuitively, this expression reminds us of Amdahl's Law: the overall speedup from more channels depends on cost of random accesses relative to streaming accesses. This is a first order approximation, and does not take into other factors such as computation (which are typically small in graph algorithms), instruction throughput, or the overlap of streaming and random accesses in actual execution. However, we believe this simple model is sufficient to highlight the importance of considering the relative performance impact on both sequential and random accesses. 

\subsection{Narrower Channel Emulation - Gap Analysis}  
\label{ssec:gap_analysis}

To further examine the effects of narrower channels, we have designed a software experiment that gives a first order approximation of having a narrower channel and its effect on cache behavior. The experiment works as follows: given an adjacency array $[x,y,z]$, we insert dummy values, $\bullet$, into the adjacency arrays such that the new adjacency is $[x,\bullet^1,\bullet^2,...,\bullet^g,y,\bullet^1,\bullet^2,...,\bullet^g,z,\bullet^1,\bullet^2,...,\bullet^g]$, where $g$ refers to the number of dummy values inserted between two edges in the original input. In the execution of the algorithm these dummy values are ignored. Specifically, we have modified Ligra to jump over the dummy values without accessing them. As such, the behavior of the analytic is not affected.

We refer to $g$ as the {\bf gap} size. A gap, ($g =1$), requires fetching in twice the number of cache lines than would be required if no dummy values are added. Generally speaking, for gap size $g$ ($g > 0$) we read in $g+1$ times as much data compared to normal. From a different perspective, the dummy values can be used to represent how a narrow channel might translate to a less effectively used cache line. For example, with a cache line of size 64 bytes and using 4 byte indices, using a gap of $g=15$ means using only a single value of each cache line that is fetched. 

Fig. \ref{fig:gap-time-pr} depicts the execution time of the graph streaming phase (upper row of sub-figures) and the total execution time (bottom row of sub-figures). Gap size of $g=0$ represents the results for the original input graph without any dummy values.
We show the execution time for multiple thread counts for both DDR and MCDRAM. 
For the DDR based memory, it is quite clear that increasing the gap increases the execution time. For 16 threads, this becomes visible at gap $g=4$. For the larger thread counts it is already apparent at $g=4$ where the execution time increases by over $50\%$.


Fig. \ref{fig:gap-scaling} shows overall execution time as a function of the number of threads. The curves correspond to different gap sizes, in powers of 2. There is not much performance difference across different gap sizes at lower thread counts (as not enough memory requests are generated). However, for DDR, performance stops scaling from 64 threads onwards similar to the results in Fig. \ref{fig:knl-gapbs} and Fig. \ref{fig:knl-ligra}. 
In contrast, MCDRAM does not see the same performance difference while scaling the number of threads. The more abundant channels on the MCDRAM are better able to handle more independent memory requests, and tested configurations do not appear to reach a saturation point.  


For the sake of brevity we do not show additional plots for this experiment, yet note that the edge streaming phase is impacted by less than $10\%$ for $g \leq 4 $. 
Note in our experiments, when $g \geq 4$, the channel width becomes smaller than 16 bytes for DRAM memory. Such a width is significantly narrower than the new channel widths suggest in \cite{zhang2014half} and \cite{o2017fine} where the new narrower channels are expected to be almost 10X larger. Thus, we expect the reduction in the channel width to have low impact on the performance of edge streaming operations. 

Using our simple performance model, Eq. \ref{eq:perf}, consider the following results for $g=4$ and 256 threads, the streaming phase accounts for $\frac{S}{R+S} = 0.3$ (an increase from 0.2 for $g=0$, thus $K=1.5$) and the edge streaming accounts for $\frac{R}{R+S} = 0.7$. Increasing the number of channels by $C = 4$ while maintaining the overall total bandwidth will improve  overall performance of \PR up by $3.47\times$ according to our model. While further study is needed, we believe that this model indicates a strong need to consider additional architectural support for systems geared towards high-performance graph algorithms.









\section{Conclusions}
\vspace{-0.1cm}

In this paper, we have demonstrated that the performance of graph algorithms depends not just on latency or bandwidth but instead is correlated with the number of memory channels available in the memory subsystem. Our experiments show that as a larger number of threads are launched on both Intel's KNL processor and NVIDIA GPUs, the high bandwidth memory system outperforms DDR by a factor that is closely tied to the ratio of memory channels in their respective memory subsystems. This scalability for high-bandwidth memory systems is almost twice is high as that of the DDR based systems, even when controlling for bandwidth and other factors.  Additionally, by using gap analysis and a simple performance model, we project that systems with narrower but more numerous channels will provide additional performance benefits for continued performance scaling up to maximum thread resources. This finding is especially important as new processors and accelerators with larger numbers of cores are deployed, and the gap between threads and memory channels continues to grow.

\bibliographystyle{IEEEtranS}
\bibliography{bibfile,bader,memory,green}

\end{document}